\documentclass[lettersize,journal]{IEEEtran}
\usepackage{amsmath,amsfonts}
\usepackage{algorithmic}
\usepackage{algorithm}
\usepackage{array}
\usepackage[caption=false,font=normalsize,labelfont=sf,textfont=sf]{subfig}
\usepackage{textcomp}
\usepackage{stfloats}
\usepackage{url}
\usepackage{booktabs}
\usepackage{verbatim}
\usepackage{graphicx}
\usepackage{cite}
\usepackage{bbding}

\begin{document}

\title{AvatarBrush: Monocular Reconstruction of Gaussian Avatars with Intuitive Local Editing}

\author{Mengtian~Li,
    Shengxiang~Yao,
    Yichen Pan,
    Haiyao~Xiao,
    Zhongmei Li,~\IEEEmembership{Member,~IEEE,}
    Zhifeng~Xie*,
    Keyu~Chen
    \IEEEcompsocitemizethanks{
\IEEEcompsocthanksitem M. Li, S. Yao, Y. Pan, and Z. Xie are with Shanghai University, Shanghai, 200444, China. 
(e-mail: {mtli, yaosx033,zhifengxie}@shu.edu.cn pany7066@gmail.com) (*Zhifeng Xie is the corresponding author.)
\IEEEcompsocthanksitem M. Li and Z. Xie are also with the Shanghai Film Visual Effects Engineering and Technology Research Center, Shanghai, 200444, China.
\IEEEcompsocthanksitem H. Xiao is with Tavus Inc., San Francisco, CA 94105, USA. (e-mail: haiyao@tavus.dev)
\IEEEcompsocthanksitem Z. Li is with the East China University of Science and Technology (ECUST), Shanghai, 200237, China. (e-mail: zhongmeili@ecust.edu.cn)
\IEEEcompsocthanksitem K. Chen is with Pinch Inc., San Francisco, CA 94105, USA. (e-mail: chern9511@gmail.com)

}
}

\markboth{Journal of \LaTeX\ Class Files,~Vol.~14, No.~8, August~2021}%
{Shell \MakeLowercase{\textit{et al.}}: A Sample Article Using IEEEtran.cls for IEEE Journals}


\maketitle
\begin{abstract}
The efficient reconstruction of high-quality and intuitively editable human avatars presents a pressing challenge in the field of computer vision. Recent advancements, such as 3DGS, have demonstrated impressive reconstruction efficiency and rapid rendering speeds. However, intuitive local editing of these representations remains a significant challenge. In this work, we propose AvatarBrush, a framework that reconstructs fully animatable and locally editable avatars using only a monocular video input. We propose a three-layer model to represent the avatar and, inspired by mesh morphing techniques, design a framework to generate the Gaussian model from local information of the parametric body model.
Compared to previous methods that require scanned meshes or multi-view captures as input, our approach reduces costs and enhances editing capabilities such as body shape adjustment, local texture modification, and geometry transfer. Our experimental results demonstrate superior quality across two datasets and emphasize the enhanced, user-friendly, and localized editing capabilities of our method.
\end{abstract}

\begin{figure*}[!t]
\centering
\includegraphics[width=0.95\textwidth]{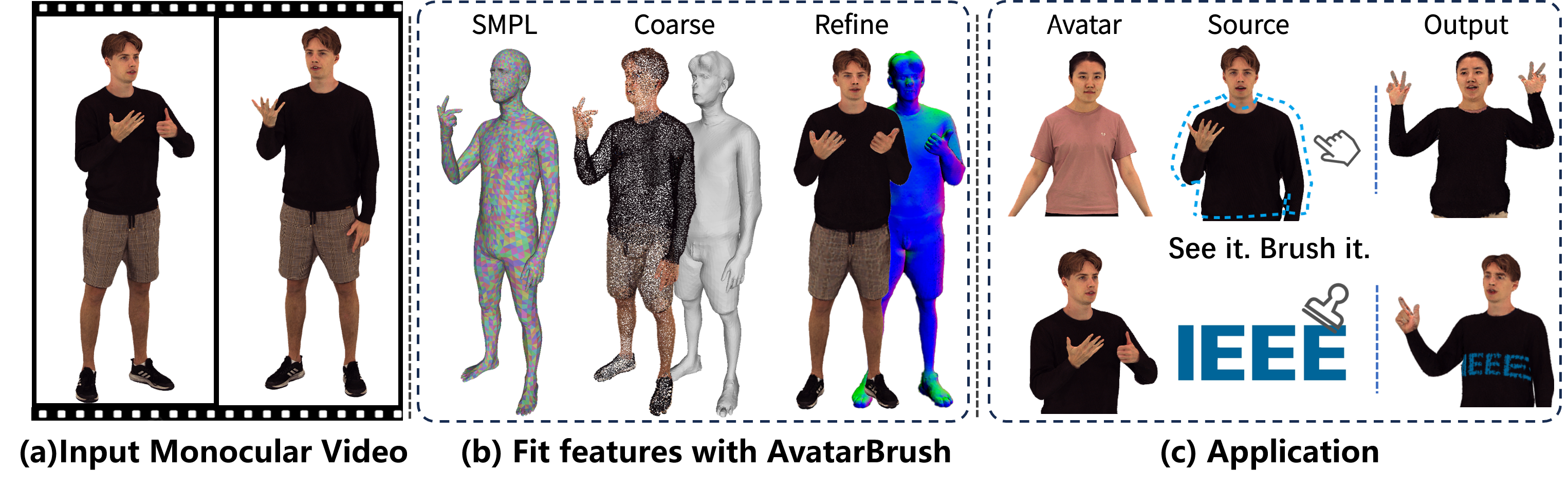}
\caption{Using a monocular video as input, we fit a set of features to generate an editable 3D avatar. Leveraging our specialized representation, GMA, this avatar can be easily edited in both texture and geometry by transferring features, while also supporting animation with hand poses and expressions. Our avatar model facilitates the transfer of garments across different identities and allows for the stamping of logos and other customizable elements in a user-friendly, interactive, real-time editing interface.}
\label{fig:teaser}
\end{figure*}
\section{Introduction}
\IEEEPARstart{C}{reating} high-fidelity clothed human models holds significant applications in virtual reality, telepresence, and movie production. Explicit methods~\cite{ref30,ref2} are generally easier to edit, facilitating straightforward adjustments to the model’s features; however, they often struggle to reconstruct high-frequency detail information directly. In contrast, implicit methods such as occupancy fields~\cite{ref42,ref43}, signed distance fields (SDF)~\cite{ref51}, and neural radiance fields (NeRFs)~\cite{ref36,ref49,ref24,ref15} have been developed to learn the clothed human body using volume rendering techniques. These implicit methods can achieve high-precision reconstruction results, but they are typically difficult to edit. Many approaches combine implicit methods with explicit methods, seeking to leverage the strengths of both techniques.

CustomHuman~\cite{ref10} combines the SMPL-X model with the SDF field to generate the human avatar, defining the geometry and texture features on the vertices of a deformable body model, stored in a codebook to exploit its consistent topology during movement and deformation. This method enables feature transfer for model editing, but it requires high-accuracy scanning data for fitting, making the construction of a new digital human very costly.

\begin{table}[t]
\caption{Comparison With Recent Works\label{table:compare_method}}
\centering
\footnotesize
\resizebox{\columnwidth}{!}{
   \begin{tabular}{lccccc}
       \hline
       Method & Monocular & Shape & Attr. Trans & Local Edit & Fast\\     
       \hline
       TAVA\hfill\cite{ref23} & \XSolidBrush  & \XSolidBrush & \XSolidBrush & \CheckmarkBold & \XSolidBrush\\
       INGP\hfill\cite{ref35} & \CheckmarkBold  & \CheckmarkBold & \XSolidBrush & \XSolidBrush & \CheckmarkBold \\
       3DGS-Avatar\hfill\cite{ref22} & \CheckmarkBold  & \XSolidBrush & \XSolidBrush & \XSolidBrush & \CheckmarkBold\\
       Splat.~Avatar\hfill\cite{ref44} & \CheckmarkBold  & \XSolidBrush & \XSolidBrush & \XSolidBrush & \CheckmarkBold\\
       Cos.~Human\hfill\cite{ref10} & \XSolidBrush  & \XSolidBrush & \CheckmarkBold & \CheckmarkBold & \XSolidBrush\\
       NECA\hfill\cite{ref50} &  \CheckmarkBold  & \CheckmarkBold  & \CheckmarkBold & \CheckmarkBold & \XSolidBrush\\
       AvatarBrush (ours) & \CheckmarkBold  & \CheckmarkBold & \CheckmarkBold & \CheckmarkBold & \CheckmarkBold \\
       \hline
   \end{tabular}
}
\end{table}

NECA~\cite{ref50} binds local features to vertices and uses tangent space to propagate these features across sampled viewpoints. This method allows neural scene modification by adjusting SMPL shape parameters and transferring clothing features via network transfer. However, NeRF-based methods have limitations in achieving finer local edits and tend to operate at slower rendering speeds. IntrinsicNGP~\cite{ref55} converts viewpoint coordinates from world space to local coordinates expressed in a UVD grid, allowing points to be represented by a hash grid, which accelerates rendering but restricts altering the model’s shape.

Our objective is to advance beyond current approaches to create realistic avatars from monocular videos, enabling efficient local editing on texture and geometry and greater computational efficiency. We observe that Gaussian-based methods~\cite{ref32,ref4,ref8} can generate models efficiently by representing the entire model as a basic point cloud, with each point depicted as a disk. To manage the point cloud, methods like GaussianAvatars~\cite{ref40} and SplattingAvatar~\cite{ref44} employ a mesh as the foundational model and set Gaussians within the local coordinate system to construct a radiance field over the mesh. This ensures a high-fidelity avatar at a low deformation cost. Zhan et al.~\cite{ref53} propose a method that leverages both the pose vector and local features derived from local anchor points on the SMPL model, to generate subtle wrinkles. However, these methods cannot directly edit avatars through Gaussian manipulation or SMPL parameter adjustment.

Modeling an editable avatar using Gaussian Splatting presents several critical challenges. One significant challenge is how to bind Gaussians to the explicit mesh while preserving editing capabilities. Simply binding Gaussians to each face and altering the SMPL surface can lead to artifacts due to the inherent differences in the properties of the Gaussian representation. To address this issue, we propose a novel representation called Gaussian Morphing Avatar (GMA), combining three layers—feature layer, mesh layer, and Gaussian layer—to create an editable avatar, as shown in Fig.~\ref{fig:framework}. We apply face offsets to generate a Gaussian-embedded mesh that preserves the original topology of the explicit parametric model. To improve the quality of our representation, we further subdivide coarse Gaussians, enhancing texture detail and capturing finer surface.

Another challenge is how to map the features to the final Gaussian. Inspired by the morphing process, which extracts features from a mesh and establishes relationships by mapping between two meshes, we introduce the Avatar Morphing. This approach assumes a target Gaussian avatar with a predefined correspondence to the explicit parametric body model. By defining Gaussian features in advance, we map each Gaussian point to the faces of the parametric model using small MLPs, establishing a static mapping from diverse feature inputs to the geometry and texture of the Gaussian representation. Through this process, we can create our three-layer GMA model. Overall, we propose a novel method to generate a locally editable clothed avatar from monocular video. Based on an explicit parametric model, the reconstructed avatar offers the animator control over body shape, pose, and hand articulation. As a benefit of our representation, users can generate novel avatars efficiently and transfer partial features or interactively paint on the model, allowing them to directly edit the model based on what they see.

\section{Related Works}
\label{sec:relatedwork}
\noindent \textbf{Editable Avatar Reconstruction.}
The primary task in reconstructing an editable avatar lies in designing an appropriate representation. TAVA~\cite{ref23} constructs a template-free volumetric actor by tracking skeletal movement and maintaining the correspondence of volume points across different poses. This correspondence ensures that the texture of the volumetric actor remains consistent across various poses. SA-NeRF~\cite{ref52} integrates sample points with the SMPL surface. NECA~\cite{ref50} maintains latent codes on the surface to apart geometry and texture, enabling relighting and localized shadow editing. But such methods consume a lot of time to train the network, and the editing processes are difficult to apply in practical scenarios.

To make the editing process more convincing and user-friendly, some work has focused on transferring the appearance attributes of avatars with different presentations to learn clothing or hair properties from RGB inputs. DELTA~\cite{ref6} combines implicit NeRF-based garment and explicit mesh-based body modeling to better represent each part. Similarly, MeGA~\cite{ref48} combines the canonical Gaussian Hair and Gaussian map-based face. GALA~\cite{ref19} utilizes DMTet~\cite{ref46} to represent the layered geometry of clothed humans from a single 3D scan. To tackle the same problem under the more challenging condition of monocular video input, $D^3$-Human~\cite{ref3} employs an SDF that acts as a strong geometric prior constraint on the solution space, allowing for the generation of coherent body and cloth surfaces via DMTet. LayGA~\cite{ref28} uses a two-layer Gaussian map to enhance the detail of cloth and enable the cloth transfer. They rely on the segmentation of appearance attributes, and most of them can't do more local editing.

CustomHuman~\cite{ref10} stores the features on the mesh vertex and creates a codebook to allow easy transfer between different identities. $E^3$Gen~\cite{ref54} uses the diffusion model to generate the Gaussian map on the SMPL UV plane and to form a shell on top of the SMPL model. But they require high-accuracy scanning data or multi-view pictures for fitting, making the construction of a new digital human costly.

Compared to other methods, we design a novel presentation to train with monocular video, and the avatar can be modified in a cost-effective and user-friendly manner to transfer part features, enabling local texture and geometry editing and body shape changing. 

\noindent \textbf{Animatable Avatar Reconstruction.}
To model an animatable avatar, many methods~\cite{ref55,ref14} use parametric models to handle dynamic scenes and obtain animatable 3D human models. Other extensions of NeRF~\cite{ref31} into dynamic scenes~\cite{ref39,ref33,ref34} and methods for animatable 3D human models in multi-view scenarios~\cite{ref27,ref36,ref24} or monocular videos~\cite{ref38,ref5,ref17} have shown promising results. Signal Distance Function (SDF) is also employed~\cite{ref26,ref16,ref7} to establish a differentiable rendering framework or use NeRF-based volume rendering to estimate the surface. Most of these methods are limited by ray sampling techniques, which restrict their rendering speed.
 
3D Gaussian Splatting (3D-GS)~\cite{ref18} is deemed as a promising improvement of the previous implicit representations. Methods~\cite{ref22,ref12,ref21} use the learnable skinning weight to associate the Gaussian point cloud to the bone transformation. Based on such an explicit method, methods~\cite{ref11,ref25,ref4,ref8} employ the Gaussian map as the pose feature to generate a Gaussian point cloud on the SMPL or SDF model. But such pose features contain both pose information and texture information that could not be edited on the UV plane. SplattingAvatar~\cite{ref44} and GaussianAvatars~\cite{ref40} both embed the Gaussian point on the mesh face to generate an animatable avatar that can be controlled with great precision using an explicit mesh. To use the SMPL-X mesh as a geometry prior, ExAvatar~\cite{ref32} encodes the vertices of the SMPL-X model to the triplane space and uses Gaussians as the presentation of the texture. To generate subtle, pose-dependent clothing wrinkles, Zhan et al.~\cite{ref53} propose a method that leverages both the global pose vector and features derived from local anchor points on the SMPL model's surface. In order to achieve efficient rendering and well-being animatable control, while decoupling part geometry information from the Gaussian point cloud, we embedded the Gaussian on the mesh and generated it by discrete local features on the SMPL-X faces.

\section{Methods}
Our goal is to reconstruct photorealistic, locally editable human avatars from monocular video inputs. Traditional NeRF-based methods~\cite{ref10,ref11} offer impressive rendering fidelity but suffer from limited interpretability and high computational costs during animation. On the other hand, parametric mesh models such as SMPL-X~\cite{ref15} provide structured control yet lack detailed surface appearance.

To bridge this gap, we propose the Gaussian Morphing Avatar (GMA), which unifies mesh-based geometry and Gaussian-based appearance within a single morphable framework. This hybrid representation maintains the kinematic consistency of parametric models while leveraging the differentiable rendering efficiency of Gaussian Splatting~\cite{ref20}. By jointly learning correspondence and deformation between these two spaces, GMA enables accurate reconstruction, fine-grained local editing, and robust re-animation under arbitrary poses.

\begin{figure*}[ht]
       \centering
       \includegraphics[width=1\linewidth]{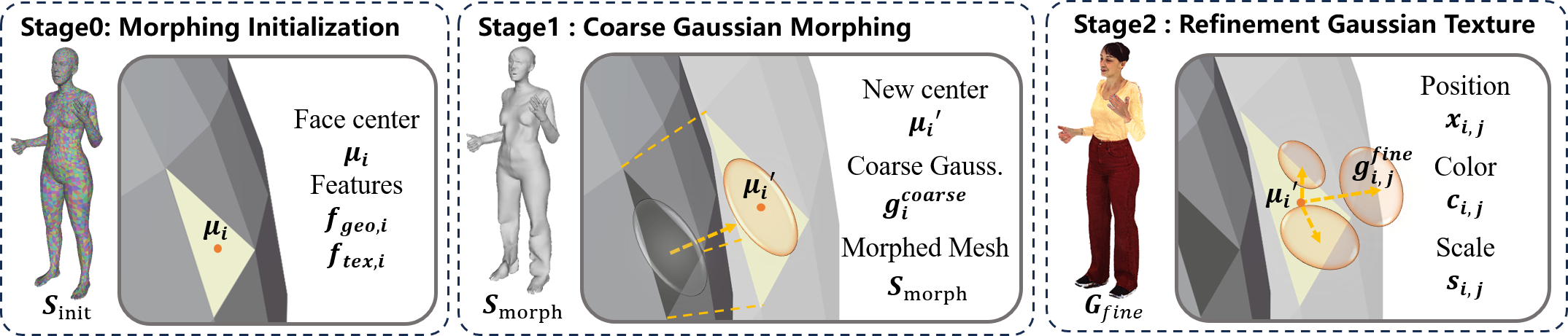}
       \caption{\textbf{The framework of Avatar morphing.}  Starting with a standard SMPL-X mesh $\mathcal{S}$, our method first learns per-face features to generate a morphed mesh $\mathcal{S}_{\text{morph}}$ that captures the coarse geometry, which is morphed by the optimization of a coarse Gaussian $\mathcal{G}_{\text{coarse}}$. This morphed mesh then serves as a scaffold to place a dense set of fine-grained Gaussians $\mathcal{G}_{\text{fine}}$, which render the final detailed appearance. The resulting layered GMA representation allows for intuitive local editing and animation.}
       \label{fig:framework}
\end{figure*}

An overview of the pipeline is shown in Fig.~\ref{fig:framework}. The input video sequence is first processed to extract per-frame SMPL-X parameters $(\beta, \theta, \psi)$, yielding a dynamic mesh sequence $\mathcal{S}_t$. We then construct a Gaussian field $\mathcal{G}_t$ anchored to this mesh via the proposed \textit{Avatar Morphing} module. The resulting representation supports both photometric supervision and geometry regularization during training, ultimately leading to a controllable, editable avatar model

\subsection{Avatar Morphing}
\noindent\textbf{Define.} We define avatar morphing as a learned mapping that generates a Gaussian-based avatar $\mathcal{G}$ whose geometry and appearance are intrinsically bound to an SMPL-X body model $\mathcal{S}$. The key to this process is establishing a dense correspondence between the SMPL-X faces and the Gaussians.

Given the input video frames $\mathcal{I}$, we first estimate the shape parameters $\beta$, pose parameters $\theta$, and expression parameters $\psi$. These parameters define the posed SMPL-X mesh, denoted as $\mathcal{S}(\beta, \theta, \psi)$, which is comprised of a dynamic vertex set $\mathbf{V}$ and the face topology $\mathbf{F} \in \mathbb{Z}^{N_f\times3}$.

The final appearance of Gaussian Morphing Avatar is represented by a set of $N_g$ Gaussians, $\mathcal{G} = \{g_n\}_{n=1}^{N_g}$. For brevity, we denote each Gaussian $g_n(\mathbf{x}_n, \mathbf{s}_n, \mathbf{c}_n)$ with its 3D position $\mathbf{x}_n$, 2D scale $\mathbf{s}_n$, RGB color $\mathbf{c}_n$.

To encode personalized geometry and texture details of the Gaussian-based avatar, we associate each face $f \in F$ of the SMPL-X mesh with a geometry feature $\mathbf{f}_{geo}\in R^{N_f\times k}$ and a texture feature $\mathbf{f}_{tex} \in R^{N_f\times k}$.

Our goal is to learn a function $\Phi$ that maps the body model state and face features to the final Gaussian parameters:
\begin{equation}
    \mathcal{G} = \Phi(\mathcal{S}, \mathbf{f}_{geo},\mathbf{f}_{geo}).
\end{equation}

\noindent\textbf{Coarse Gaussian Morphing.} In this initial stage, we construct a coarse geometric scaffold by associating a single Gaussian $g_i^{\text{coarse}}(\mathbf{x}_i^{\text{init}})$ with each face $f_i$ of the SMPL-X mesh $\mathcal{S}$. We use an MLP $\mathcal{F}_{\text{coarse}}$ to predict the final position for each coarse Gaussian. The network takes the geometry feature of that specific face, $\mathbf{f}_{\text{geo}, i}$, and its face center after position encoding~\cite{ref31}, $\hat\mu_i$, as input:
\begin{equation}
    \mathbf{x}_i^{\text{final}} = \mathcal{F}_{\text{coarse}}(\mathbf{f}_{\text{geo}, i}, \hat\mu_i).
    \label{eq:coarse_pos}
\end{equation}

A key principle of our method is that each coarse Gaussian, $g_i^{\text{coarse}}$, is anchored to the center $\boldsymbol{\mu}_i$ of its corresponding mesh face $f_i$. During the main training phase, the standard 2D-GS optimizes all Gaussian parameters, including their positions, to reconstruct the target appearance from the training images. As the initial positions $\mathbf{x}_i^{\text{init}}$ are optimized to their final positions $\mathbf{x}_i^{\text{final}}$, they effectively "pull" the underlying mesh structure along with them.

This optimization-driven movement morphs the base SMPL-X mesh into a morphed mesh. We define the resulting morphed mesh $\mathcal{S}_{\text{morph}}$ by considering the final Gaussian positions $\mathbf{x}_i^{\text{final}}$ as the new target centers $\mu_i'$ for the mesh faces with the final face offset $O$.


\noindent\textbf{Fine-grained Gaussian Generation.} The coarse Gaussians established in the previous stage provide a foundational geometric scaffold but cannot represent fine details such as wrinkles, hair strands, or clothing folds require higher spatial resolution. To enhance the model's expressiveness, we generate a set of fine-grained Gaussians, $\mathcal{G}_{\text{fine}}$, anchored to the morphed mesh $\mathcal{S}_{\text{morph}}$.

Specifically, for each face $f_i$ of the offset mesh, we generate $N_k$ new Gaussians, indexed by $j=1, \dots, N_k$. The position of each new Gaussian, $g_{i,j}$, is determined by a second MLP, $\mathcal{F}_{\text{fine}}$. This network learns to map the geometry feature $\mathbf{f}_{\text{geo}, i}$ and face center after position encoding $\hat\mu_i$. This allows for a detailed distribution of Gaussians across the surface:
\begin{equation}
    \mathbf{x}_{i,j} = \mathcal{F}_{\text{fine}}(\mathbf{f}_{\text{geo}, i}, \hat\mu_i).
    \label{eq:fine_pos}
\end{equation}

\noindent\textbf{Gaussian Appearance.} In addition to geometry, we decode color $\mathbf{c}$ and scale $\mathbf{s}$ for each fine-grained Gaussian $g_{i,j}$. These properties are derived from the corresponding face's texture feature, $\mathbf{f}_{\text{tex}, i}$. We use two separate decoder networks, $\mathcal{T}_{\text{color}}$ and $\mathcal{T}_{\text{scale}}$, which also take the face center $\hat\mu_i$ as input to embed the spatial information within a single face:
\begin{equation}
\begin{split}  
    \mathbf{c}_{i,j} &= \mathcal{T}_{\text{color}}(\mathbf{f}_{\text{tex}, i}, \hat\mu_i) \\
    \mathbf{s}_{i,j} &= \mathcal{T}_{\text{scale}}(\mathbf{f}_{\text{tex}, i}, \hat\mu_i)~~.
    \label{eq:texture_decode}
\end{split}
\end{equation}

Ultimately, our \textbf{Gaussian Morphing Avatar (GMA)} is a cohesive, three-layer representation. It consists of: the feature layer on the parametric model $\mathcal{S}$; the morphed Mesh $\mathcal{S}_{\text{morph}}$ with its detailed geometry and the dense set of fine-grained Gaussians $\mathcal{G}_{\text{fine}}$ that render the final, photorealistic appearance. This structure provides a robust and flexible model for both animation and high-fidelity local editing.

\subsection{Training Strategy}

%
%
\noindent \textbf{Two-stage training.}
In the design of the hybrid representation, we need to record the texture and geometry information of an avatar. It is important to fit the different information into two stages. Like the AnimatableGaussian~\cite{ref25}, they use an off-shell method to generate an SDF as the geometry prior. We split the geometry and texture fitting into two stages. The first stage focuses on the geometry of the model, emphasizing the importance of geometric accuracy over texture information. We assign greater weight to the geometry loss during this stage compared to the second stage. In the second stage, we freeze the coarse geometry MLP $\mathcal{F}_{coarse}$ and the geometry features $\mathbf{f}_{geo}$ to maintain the results from the first stage while reducing the weights of the geometry loss.

\noindent \textbf{Geometry loss.} To fit a mesh through Gaussian splatting from monocular video, due to the gap between the mesh surface and the gaussians, the model might be fitted in the image space, but it ignores the real geometry. Follow SoftRas~\cite{ref29}, we use Laplacian loss to restrict the model $ L_{lap} = \sum_{i} \| L \cdot V_i \|^2$. To gain a smoother mesh, we use the relative edge loss $L_{edge}$ from~\cite{ref9} between the optimized body mesh with and without applied offsets. We also use the normal loss from 2D-GS~\cite{ref13} to restrict the surface. Above all, our geometry loss could be present as:
\begin{equation}
    L_{geo} = \lambda_{lap}L_{lap} + \lambda_{edge}L_{edge} + \lambda_{normal}L_{normal}, 
    \label{eq:geo_loss}
\end{equation}
\textbf{Picture loss.} Based on the Gaussian splatting, we use the L1 distance loss $L_{L1}$ and SSIM loss $L_{ssim}$. To decrease the float Gaussian, we use the mask loss $L_m$. In the second stage to gain better texture details, we use the lpips loss $L_{lpips}$. Above all, our picture loss could be present as:
\begin{equation}
    L_{pic} = \lambda_{L1}L_{L1} + \lambda_{ssim}L_{ssim} + \lambda_{m}L_{m} + \lambda_{lpips}L_{lpips}, 
    \label{eq:pic_loss}
\end{equation}
The picture loss and geometry loss would have different weights in two stages to focus on geometry or texture.

\begin{figure}
        \centering
        \includegraphics[width=1\linewidth]{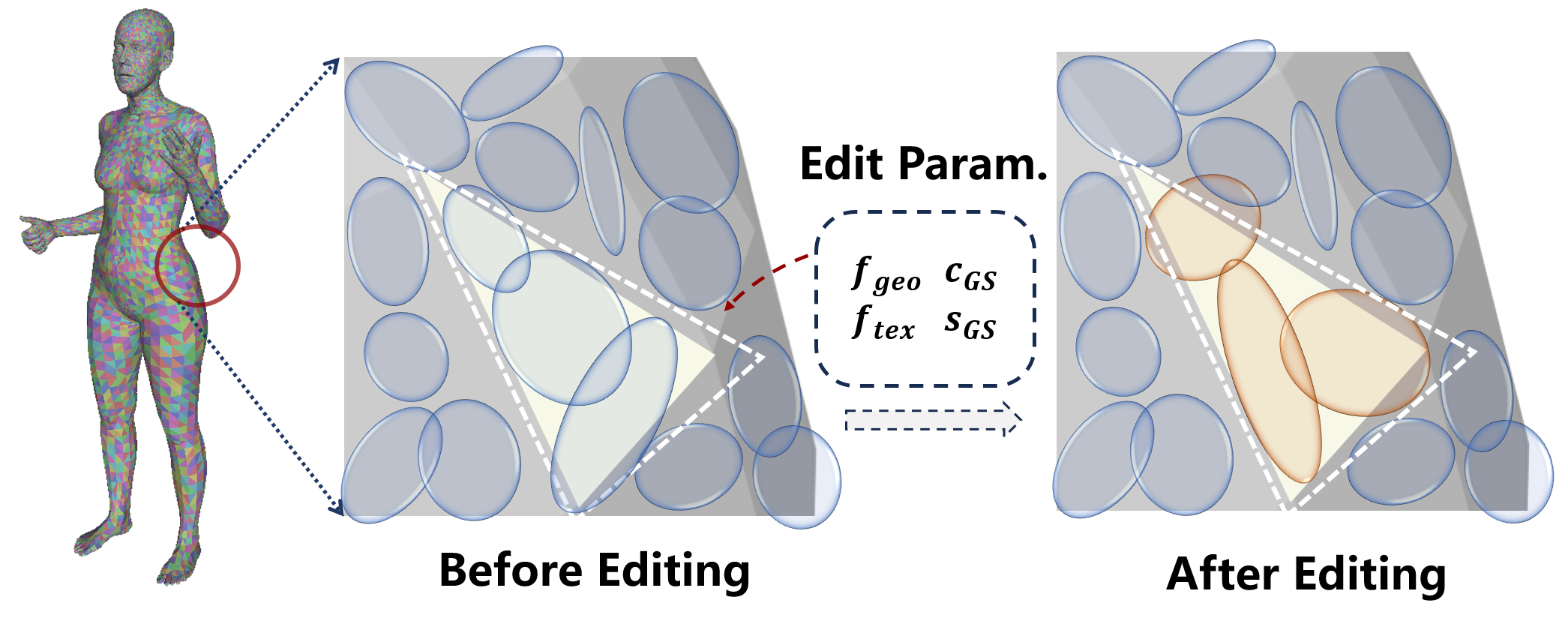}
        \caption{\textbf{Direct edit with feature.} Users can select specific faces through the interface to modify the corresponding features for localized editing on texture or geometry.}
        \label{fig:Edit Method}
\end{figure}


\subsection{Interactive Edit}
\noindent\textbf{Texture and Appearance Editing.}
Users can edit the avatar's appearance by either directly painting colors on the surface or projecting an image-based pattern. For the edited Gaussian, this defines a target color $\mathbf{c}_{\text{edit}}$. To make the edit persistent, we update the corresponding texture features $\{\mathbf{f}_{\text{tex}, i}\}$ by inverting our frozen appearance decoder $\mathcal{T}_{\text{color}}$ to find the features that best reproduce the target colors $\mathbf{c}_{\text{edit}}$.

\noindent\textbf{Cross-Identity Feature Transfer.}
Our representation allows for transferring geometric or appearance characteristics between two different avatars, a source $A$ and a target $B$. After a user selects a corresponding region on both avatars (e.g., the upper body), the feature vectors of the selected faces on avatar $B$ are replaced with those from avatar $A$. This transfer can be applied to geometry ($\mathbf{f}_{\text{geo}}$), texture ($\mathbf{f}_{\text{tex}}$), or both. This decoupled nature enables powerful edits, such as applying the texture of a denim jacket from one avatar onto the geometry of a hoodie on another.

\noindent\textbf{Parametric Shape and Pose Control.}
Users can perform body shape modifications by adjusting the shape parameters $\beta$ and create full-body animations by manipulating the pose parameters $\theta$ and the expression parameters $\psi$.

\begin{table*}[!t]
\caption{\textbf{Quantitative comparison on Xhuman dataset.} 
Our approach demonstrates a significant advantage in metric comparisons, particularly in terms of LPIPS, which reflects the recognition of humans. Additionally, our method supports high-quality reconstruction and provides editability, allowing for precise modifications while maintaining detail and accuracy.}
\label{table:Xhuman_compare}
\centering
\resizebox{\textwidth}{!}{   
    \begin{tabular}{lcccccccccccc}
        \toprule
                    &\multicolumn{3}{c}{00016 (male)} & \multicolumn{3}{c}{00019(female)} & \multicolumn{3}{c}{00018 (male)} & \multicolumn{3}{c}{00027 (female)}\\
                    &\textbf{PSNR↑} & \textbf{SSIM↑} & \textbf{LPIPS↓}  & \textbf{PSNR↑} & \textbf{SSIM↑} & \textbf{LPIPS↓} & \textbf{PSNR↑} & \textbf{SSIM↑} & \textbf{LPIPS↓} & \textbf{PSNR↑} & \textbf{SSIM↑} & \textbf{LPIPS↓} \\
        \midrule
             GART\hfill\cite{ref22}   &25.71 &0.9295 &0.0598 &27.78 &0.9512 &0.0668 &30.98 &0.9595 &0.0683& 26.56 &0.9449 &0.0595\\
             3DGS-Avatar\hfill\cite{ref41} &  25.44 &0.9315 &0.0409  &27.63 &0.9539 &0.0471  &28.71 &0.9521 &0.0580&26.84 &0.9477 &0.0445 \\
             \midrule
              HAHA\hfill\cite{ref47}   & 25.49 &0.9339 &0.0507 &28.49 &0.9593 &0.0501 &31.10 &0.9630 &0.0579 &27.26 &\textbf{0.9513} &0.0473\\
             SplattingAvatar\hfill\cite{ref44} &\textbf{26.73}&0.9316&0.0888&\textbf{29.28}&0.9597&0.0684&\textbf{31.89}&0.9637&0.0790&\textbf{27.30}&0.9483&0.0755\\
             \textbf{Ours}\hfill &26.10&\textbf{0.9359}&\textbf{0.0435}&28.70 &\textbf{0.9631}&\textbf{0.0442}&31.56&\textbf{0.9658}&\textbf{0.0478}&26.95&0.9493&\textbf{0.0458}\\
        \bottomrule
    \end{tabular}}
\end{table*}

\begin{table*}[!t]
\caption{\textbf{Quantitative comparison on ZJU\_MoCap dataset.} 
Compared with the result of training with the original dataset parameters, we gain a comparable result. Furthermore, in our comparison with SplattingAvatar~\cite{ref44}, which utilizes the same SMPL-X parameters as input, our approach consistently achieves favorable scores across the key evaluation metrics.}
\label{table:ZJU_compare}
\centering
\resizebox{\textwidth}{!}{   
    \begin{tabular}{lcccccccccccccccccc}
        \toprule
                    &\multicolumn{3}{c}{377} &\multicolumn{3}{c}{386} & \multicolumn{3}{c}{387} & \multicolumn{3}{c}{392} & \multicolumn{3}{c}{393} & \multicolumn{3}{c}{394}\\
                    &\textbf{PSNR↑} & \textbf{SSIM↑} & \textbf{LPIPS↓} 
                    &\textbf{PSNR↑} & \textbf{SSIM↑} & \textbf{LPIPS↓}  
                    &\textbf{PSNR↑} & \textbf{SSIM↑} & \textbf{LPIPS↓} 
                    &\textbf{PSNR↑} & \textbf{SSIM↑} & \textbf{LPIPS↓} 
                    &\textbf{PSNR↑} & \textbf{SSIM↑} & \textbf{LPIPS↓} 
                    &\textbf{PSNR↑} & \textbf{SSIM↑} & \textbf{LPIPS↓} \\
        \midrule
             GART\hfill\cite{ref22}   &34.18&0.9879&0.0157 &34.40&0.9810&0.0283 &29.34&0.9702&0.0366 &32.94&0.9787&0.0306 &30.21&0.9705&0.0356&32.33&0.9757&0.0285 \\
             3DGS-Avatar\hfill\cite{ref41}  &30.64&0.9774&0.0208 &33.63&0.9773&0.0257 &28.33&0.9642&0.0342 &31.66&0.9730&0.0301 &28.88&0.9635&0.0352 &30.54&0.9661&0.0312 \\
             NECA\hfill\cite{ref50} &25.48&0.9786&0.0233&26.95&0.9774&0.0314&22.47&0.9610&0.0432 &26.41&0.9726&0.0358 &23.44&0.9631&0.0513&24.88&0.9660&0.0374\\
        \midrule
             SplattingAvatar\hfill\cite{ref44} &29.09&0.9633&0.0467&31.66&0.9628&0.0498 &27.15&0.9489&0.0600 &29.61&0.9569&0.0550&27.62&0.9481&0.0570 &29.06&0.9513&0.0545\\
             \textbf{Ours} &30.01&0.9696&0.0247&31.77&0.9687&0.0342&27.34&0.9512&0.0417&30.02&0.9655&0.0365&28.09&0.9511&0.0384&29.46&0.9536&0.0349\\
        \bottomrule
    \end{tabular}}
\end{table*}

\begin{figure*}
        \centering
        \includegraphics[width=1\linewidth]{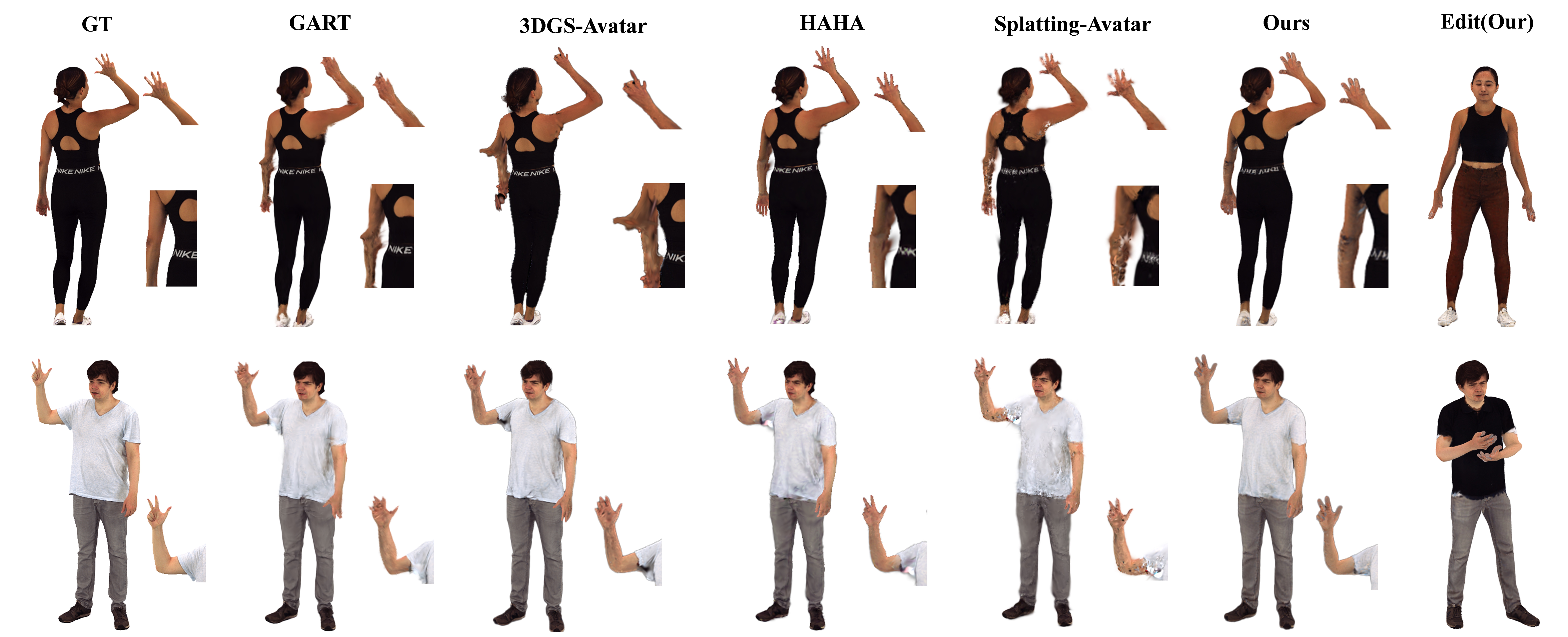}
        \caption{\textbf{Novel pose compare on X-Human.} We show the results for novel pose animation and garment transfer results. Our method produces high-quality reconstruction results and, compared to other methods, we can reduce the floating artifact in novel poses. Our method allows for editing of clothing style and texture.}
        \label{fig:xhuman compare}
\end{figure*}

\begin{figure*}
        \centering
        \includegraphics[width=1\linewidth]{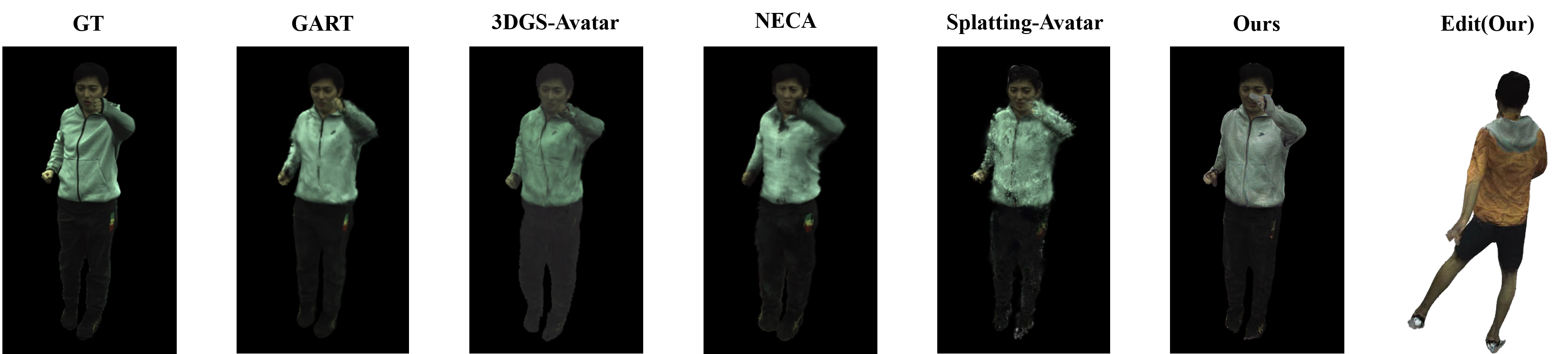}
        \caption{\textbf{Novel view compare on ZJU-mocap.} As the results for novel view synthesis and local geometry edit result. Our method ensures a robust and consistent alignment of the garment with the underlying body motion. We also demonstrate its effectiveness by transferring the hood to another avatar in a novel pose.}
        \label{fig:zju compare}
\end{figure*}


\begin{table}[!t]
\caption{\textbf{Ablation Study.} As validated by our ablation studies, the full model not only achieves superior reconstruction quality over its ablated variants, but also successfully disentangles texture from geometry.}
\label{table:as_metrics}
\centering
\small
\setlength{\abovecaptionskip}{1mm}
\begin{tabular}{lccc}
        \toprule
        Method & PSNR & SSIM & LPIPS \\      
        \midrule
        Full model & \textbf{28.32} & \textbf{0.9535} & \textbf{0.0453} \\
        One stage & 25.36 & 0.9417 & 0.0454 \\
        Mesh layer & 27.79 & 0.9505 & 0.0547 \\
        Gaussian layer & 20.40 & 0.9155 & 0.0710 \\
        \bottomrule
\end{tabular}
\end{table}

\section{Experiment And Results}
\label{sec:exp}
\subsection{Implementation Details.} 
AvatarBrush is implemented in PyTorch and optimized using the Adam~\cite{ref20} optimizer. We adopt a two-stage training strategy to achieve stable convergence and better disentanglement between geometry and texture representations. The model fitting for a novel identity involves 40k optimization steps in total, with 20k steps for each stage. In the first stage, we focus on coarse geometric reconstruction, setting the hyper-parameters as $\lambda_{Lap}=1000$, $\lambda_{edge}=100$, and $\lambda_{normal}=0.05$ to enhance mesh smoothness and preserve local details. In the second stage, we refine the texture and appearance while freezing the geometry, setting the hyper-parameters as $\lambda_{Lap}=100$, $\lambda_{edge}=10$, $\lambda_{normal}=0$, and $\lambda_{lpips}=0.02$ to balance perceptual and photometric consistency.\\  
For fitting a novel identity, the entire process takes about 1.5 hours on a single NVIDIA RTX 4090 GPU, while rendering proceeds at approximately 54 FPS, demonstrating the efficiency and real-time capability of our representation. \\

\subsection{Setup.} 
We train our model using 5 distinct subsets from the X-Human dataset, ensuring that none of them overlap with the evaluated test datasets. Each subset contains around 1000 frames covering a wide range of human poses, camera angles, and clothing styles, providing sufficient diversity for generalization. For each novel avatar reconstruction, we fit the feature parameters of the new identity while keeping the MLPs frozen to avoid overfitting.  
We compare AvatarBrush against several state-of-the-art methods, including Gart~\cite{ref22}, 3DGS-Avatar~\cite{ref41}, SplattingAvatar~\cite{ref44}, HAHA~\cite{ref47}, and NECA~\cite{ref50}. Quantitative and qualitative evaluations are conducted on two public benchmarks: the X-Human Dataset~\cite{ref45} for pose generalization and the ZJU-MoCap Dataset~\cite{ref36} for novel-view synthesis.  
To ensure fair comparison, we follow the preprocessing pipelines used in prior works, including camera calibration, background masking, and SMPL-X parameter extraction. Each model is trained under the same rendering resolution and evaluated using PSNR, SSIM, and LPIPS metrics.

\subsection{Detailed architecture of MLPs} 
\begin{figure}[hb]
    \centering
    \includegraphics[width=\linewidth]{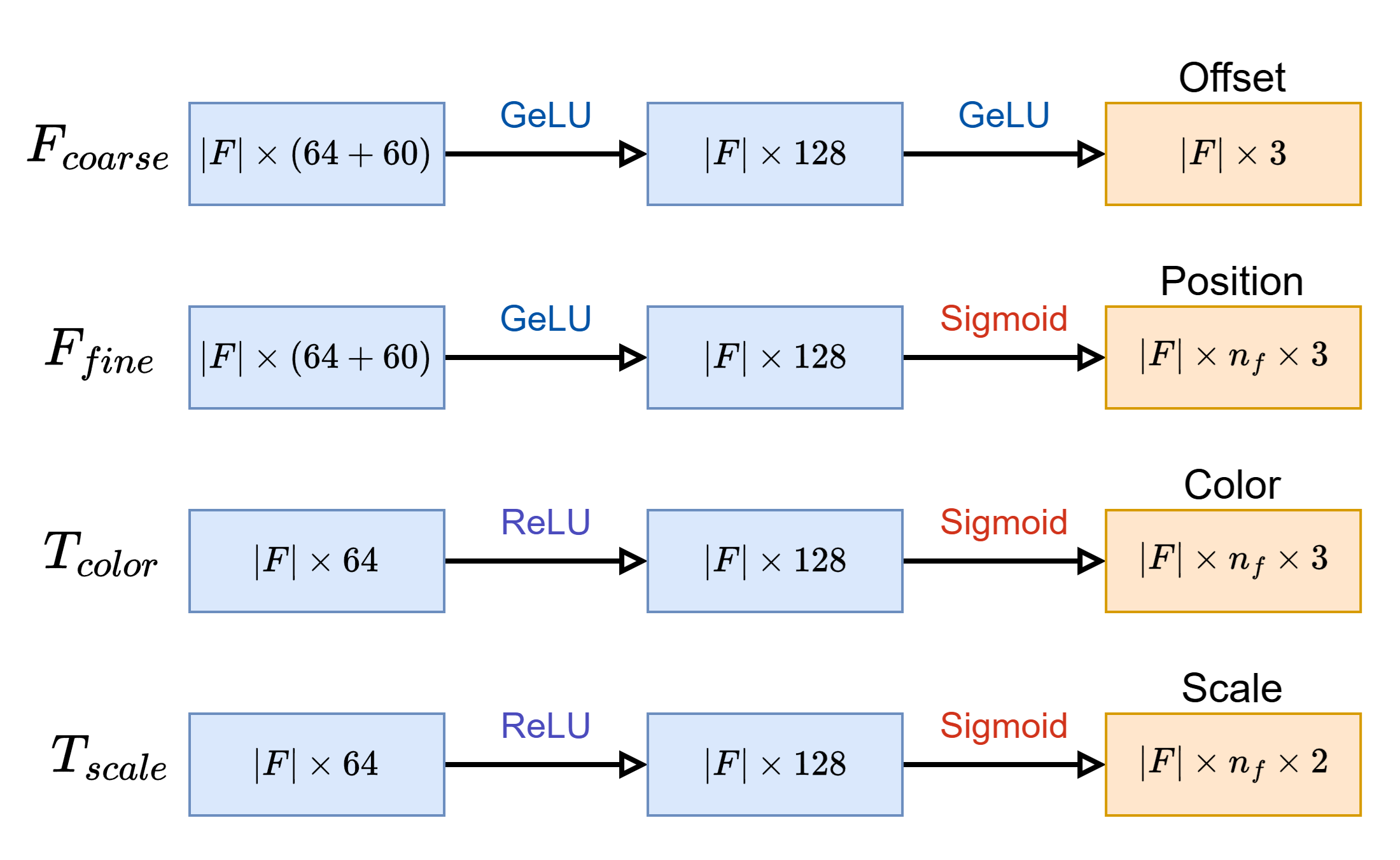}
    \caption{\textbf{MLP structures.}We design four MLPs to generate the GMA model: $F_{coarse}$ predicts the coarse geometry offset; $F_{fine}$ generates the fine-grained Gaussian positions; and $T_{color}$ and $T_{scale}$ decode the color and scale properties, respectively.}
    \label{fig:MLP_str}
\end{figure}

In the whole frame work, we design four MLPs to generate different part of the GMA model. These $F*$ are implemented in a $LINEAR \to GELU \to LINEAR$ style and these $ T*$ are implemented in a $LINEAR \to RELU \to LINEAR$ style with the hidden dimension of 128, as illustrated in Fig.~\ref{fig:MLP_str}. Each branch’s output is activated with a head layer. In the first stage, we gain the color of Gaussian by average the color of output.

\subsection{Gaussian embed on Mesh}
To embed the Gaussian, we align the gaussian with corresponding face. At first, we define the gaussian as $G_c(x_f,\hat{s},\hat{q}, c, 1)$, where $\hat{s}=(1,1)$ and $\hat{q}$ is the unit quaternion. In the first stage, After changing to $\mathcal{M}$ with:
\begin{equation}
    X_i = X_f + \frac{1}{3}\sum_k O^{scale}_k * N_{v,k},
\end{equation}
$O^{scale}$ is the vertex offset scale of the face, and $N_{v,i}$ is the normal of the vertex. We could gain the transformation of each face, such as the new position, the rotation $q_f$, and the area $s_f$. The Gaussian model would be represent as: $G(x_i, s_f*\hat{s}, q_f*\hat{q}, c, 1)$. 

In the second stage, we use the uvd coordinate to present the Gaussians, the position $X_{ij}$ of $G(x_{ij})$ could be calculated as:
\begin{equation}
    X_{ij} = u * V_1 + v * V_2 + (1-u-v) * V_3 + N_f * d,
\end{equation}
where ${V_1,V_2,V_3}$ is the position of vertices and $N_f$ is the normal on the corresponding face on the $\mathcal{S}_{\text{init}}$. Similar to the first stage, the Gaussian model would become: $G(x_i, s_f*s*\hat{s}, q_f*\hat{q}, c, 1)$.

\subsection{Experiment Result}
\textbf{Novel-pose Comparisons.} As shown in Table~\ref{table:Xhuman_compare}, our method consistently outperforms existing approaches in most evaluation metrics for novel-pose synthesis. The superior results stem from our explicit control of Gaussian rotation and translation guided by the underlying mesh structure, which enables a more physically consistent motion representation. Unlike MLP-based linear blend skinning fields, which rely on implicit deformation, our mesh-guided strategy preserves local geometric coherence and minimizes texture distortion under extreme poses.  
Qualitative comparisons in Fig.~\ref{fig:xhuman compare} reveal that methods such as Gart~\cite{ref22} and 3DGS-Avatar~\cite{ref41} often exhibit ghosting and texture tearing artifacts when reconstructing unseen poses. SplattingAvatar~\cite{ref44} and HAHA~\cite{ref47} suffer from floating Gaussian clusters around limbs and torsos, which degrade the visual realism. In contrast, our method maintains structural integrity and achieves visually smooth deformations across diverse body motions.  
Furthermore, leveraging the SMPL-X model allows our approach to generate highly realistic hand articulations and facial dynamics, supporting high-fidelity animation and precise local editing even in complex motion sequences.

\noindent\textbf{Novel-view Comparisons.} As illustrated in the Table.~\ref{table:ZJU_compare}, our method demonstrates comparable results in novel view synthesis. On the one hand, the ZJU-MoCap dataset contains numerous non-rigid transformations, which our method is unable to generate. On the other hand, we refit the pose parameters of ZJU-MoCap on SMPL-X, which introduces some noise. Compared to SplattingAvatar~\cite{ref44}, which uses the same parameters, we achieve better results across all metrics.
Fig.~\ref{fig:zju compare} shows that SplattingAvatar~\cite{ref44} generates many floating points, indicating difficulty in aligning different poses. NECA~\cite{ref50} fails to generate proper novel-view results, while 3DGS-Avatar~\cite{ref41} and Gart~\cite{ref22} exhibit missing textures on clothing. In contrast, our method maintains the garment details on the avatar, such as the zipper, whereas the other methods lose these features when the view changes. In the right of Fig.~\ref{fig:zju compare}, we could transfer the part of the garment locally to another avatar, such as the hood. More results could be found in the Appendix.

\begin{figure}
        \centering
        \includegraphics[width=1\linewidth]{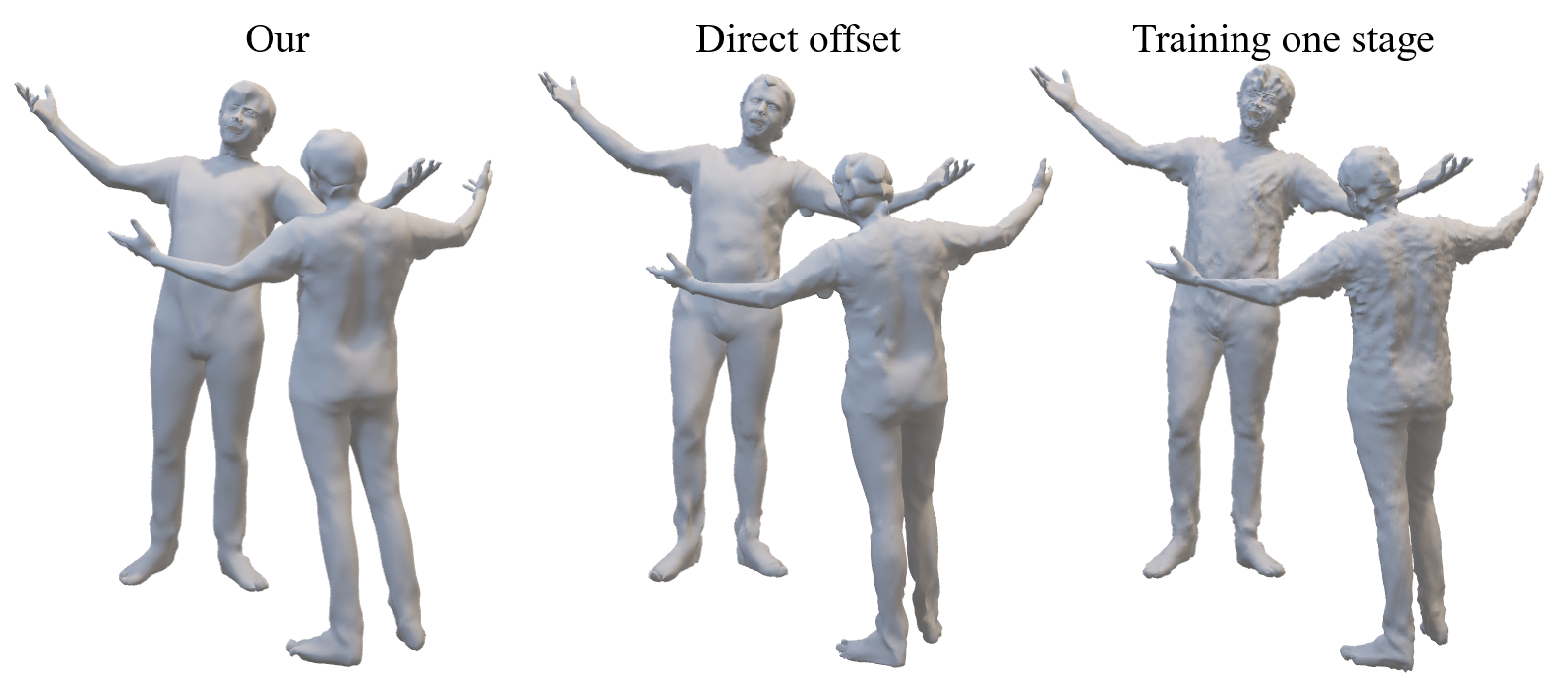}
        \caption{\textbf{Ablation study result on mesh.} Compared to directly applying face offsets, our method generates a smoother and more aligned mesh. A noisy mesh is a common artifact of single-stage training, as this approach often conflates texture information with geometric structure.}
        \label{fig:Ablation_mesh}
\end{figure}

\begin{figure}
        \centering
        \setlength{\abovecaptionskip}{1mm}
        \includegraphics[width=1\linewidth]{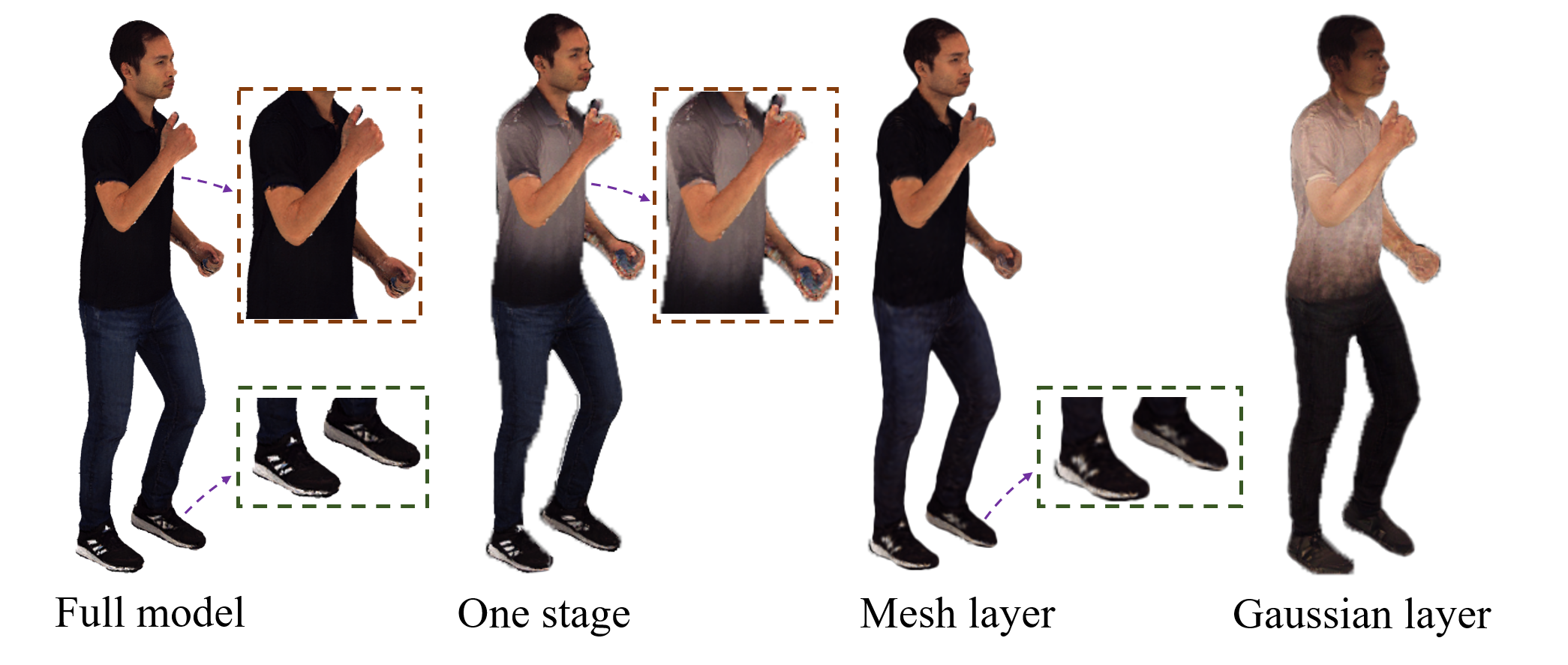}
        \caption{\textbf{Ablation study result on texture.} Single-stage training causes confusion between the texture and geometry spaces, potentially leading to incorrect texture mapping. Optimizing only a coarse set of Gaussians results in a model that lacks significant texture details, particularly in areas like the shoes. Directly embedding the Gaussian onto the local coordinate of the SMPL-X model makes it challenging to optimize the MLPs to gain proper results.}
        \label{fig:Ablation_pic}
\end{figure}

\subsection{Ablation Study}
\noindent \textbf{Three layer structure.}
To ensure editability, it is crucial to constrain the Gaussians to specific local regions on the surface. This restriction allows for more precise control over the changes and ensures that the modifications are localized to the desired areas. As illustrated in Fig.~\ref{fig:Ablation_pic}, applying the Gaussian directly to the mesh layer without properties optimization could capture most of the texture, but it struggles with finer details. Conversely, when we only focus on optimizing the properties of the embedded Gaussian points using local coordinates, the MLP becomes difficult to optimize effectively to cover sufficient detailed information.

\noindent \textbf{Two-stage Training.}
A single-stage approach to generating the final Gaussian avatar often compromises fine geometric details and hinders the creation of accurate textures. This single-step approach struggles to achieve smooth geometry, usually resulting in artifacts in the face region and reducing the overall fidelity of the model, as shown in Fig.~\ref{fig:Ablation_mesh}. Furthermore, as demonstrated in Fig.~\ref{fig:Ablation_pic}, the color MLP tends to overfit the training data, hindering its ability to generalize to the appearance of new identities.

\noindent \textbf{Mesh offset.}
To reconstruct the mesh layer, we explored two methods for calculating mesh offsets. One method generates face offsets and transfers these values to the corresponding vertices. The other calculates the normal scale for each vertex, which is then multiplied by the vertex normal to determine the offset. In Fig.~\ref{fig:Ablation_mesh}, optimizing the face offsets directly can lead to local collapses, even when applying edge loss for smoothing. This issue arises from the discrepancy between the mesh surface and the Gaussian representation, causing some Gaussians to expand excessively to display texture while neglecting the underlying surface. As a result, this approach disregards the mesh surface, leading to a loss of local alignment with the SMPL-X faces. In contrast, using normal-based offsets helps maintain a smoother mesh and preserves the original topology more effectively.

\section{Application}

\subsection{Feature Transfer.} 
\begin{figure}[ht]
        \centering
        \setlength{\abovecaptionskip}{1mm}
        \includegraphics[width=1\linewidth]{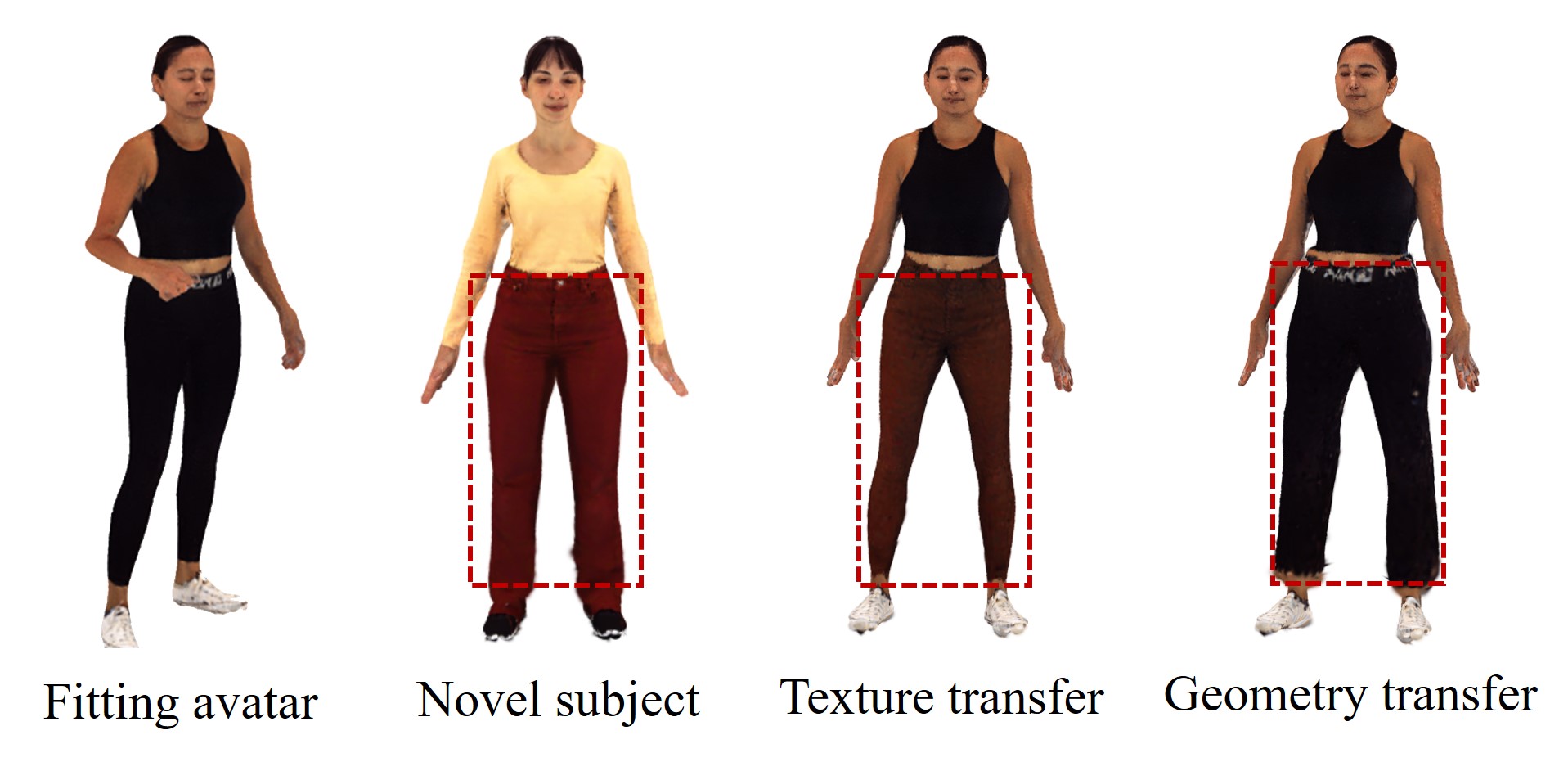}
        \caption{\textbf{Feature transfer.} Our representation effectively decouples geometry from texture, enabling independent local editing. For instance, a user can alter the shape of pants to create a wide-leg style while preserving the original fabric pattern. Conversely, the material can be changed from cotton to denim without affecting the underlying geometry.}
        \label{fig:feature transfer1}
\end{figure}

\begin{figure}[h]
        \centering
        \setlength{\abovecaptionskip}{1mm}
        \includegraphics[width=1\linewidth]{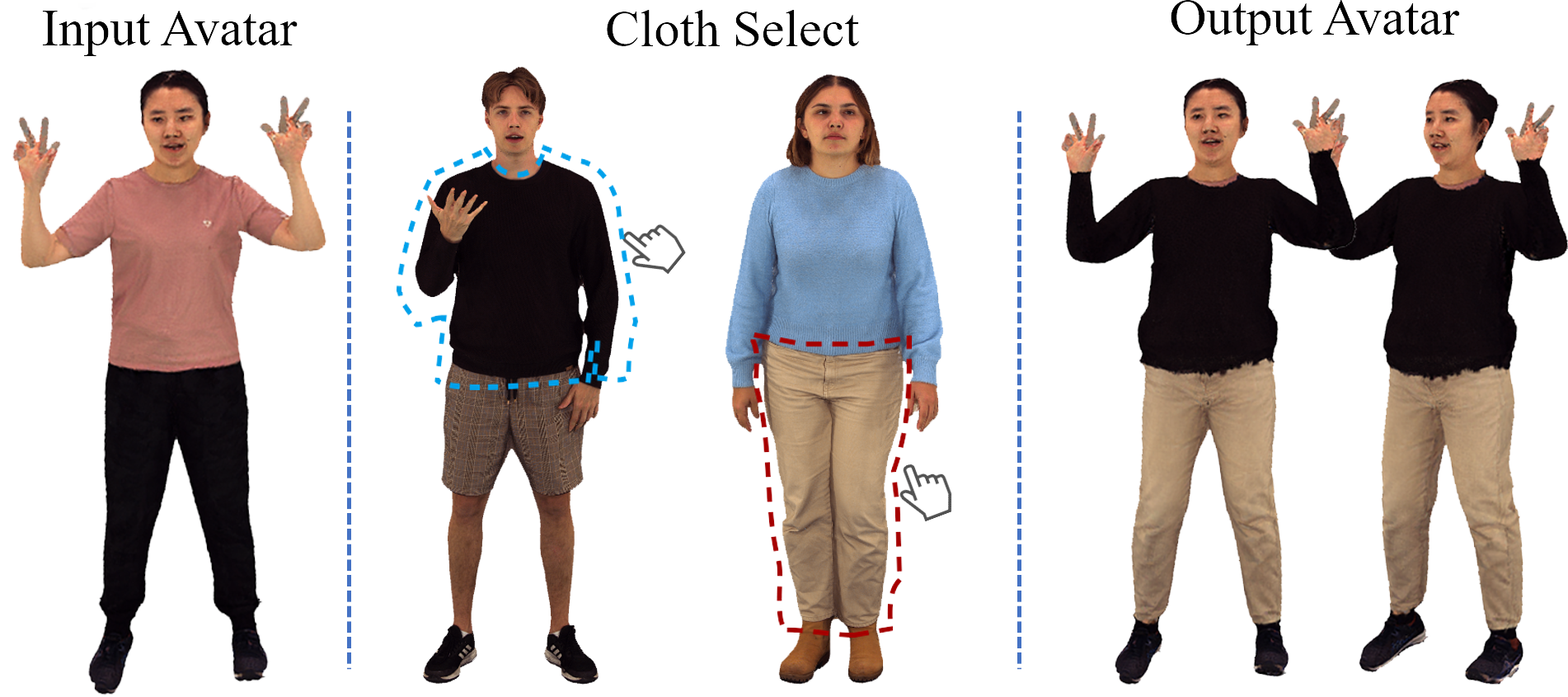}
        \caption{\textbf{Garment transfer. } By leveraging a segmentation of the clothing faces, our method can seamlessly transfer garments across different identities.}
        \label{fig:feature transfer2}
\end{figure}

Our representation naturally enables feature-level disentanglement between geometry and texture, facilitating flexible attribute manipulation. As illustrated in Fig.~\ref{fig:feature transfer1}, users can modify local parts such as pants or shirts independently, changing color, shape, or style while preserving global consistency. For example, the black yoga trousers can be converted into a brown or wide-leg design without altering the rest of the outfit. Similarly, users can add new components such as hoods or accessories to extend the avatar’s appearance. In Fig.~\ref{fig:feature transfer2}, garments are transferred seamlessly between different identities, enabling cross-character editing and virtual try-on applications. 

\subsection{Texture Stamp.} 
In addition to semantic-level editing, our system supports intuitive, brush-based texture modification. Users can directly paint on Gaussian points or stamp predefined images, such as logos or emblems, onto the avatar surface. This process requires no additional retraining and provides real-time visual feedback, allowing for rapid content personalization. As demonstrated in Fig.~\ref{fig:teaser}, this capability enhances interactivity and broadens the usability of our system in creative and entertainment scenarios.

\subsection{Animation and Shape edit.} 
\begin{figure}[h]
        \centering
        \setlength{\abovecaptionskip}{1mm}
        \includegraphics[width=1\linewidth]{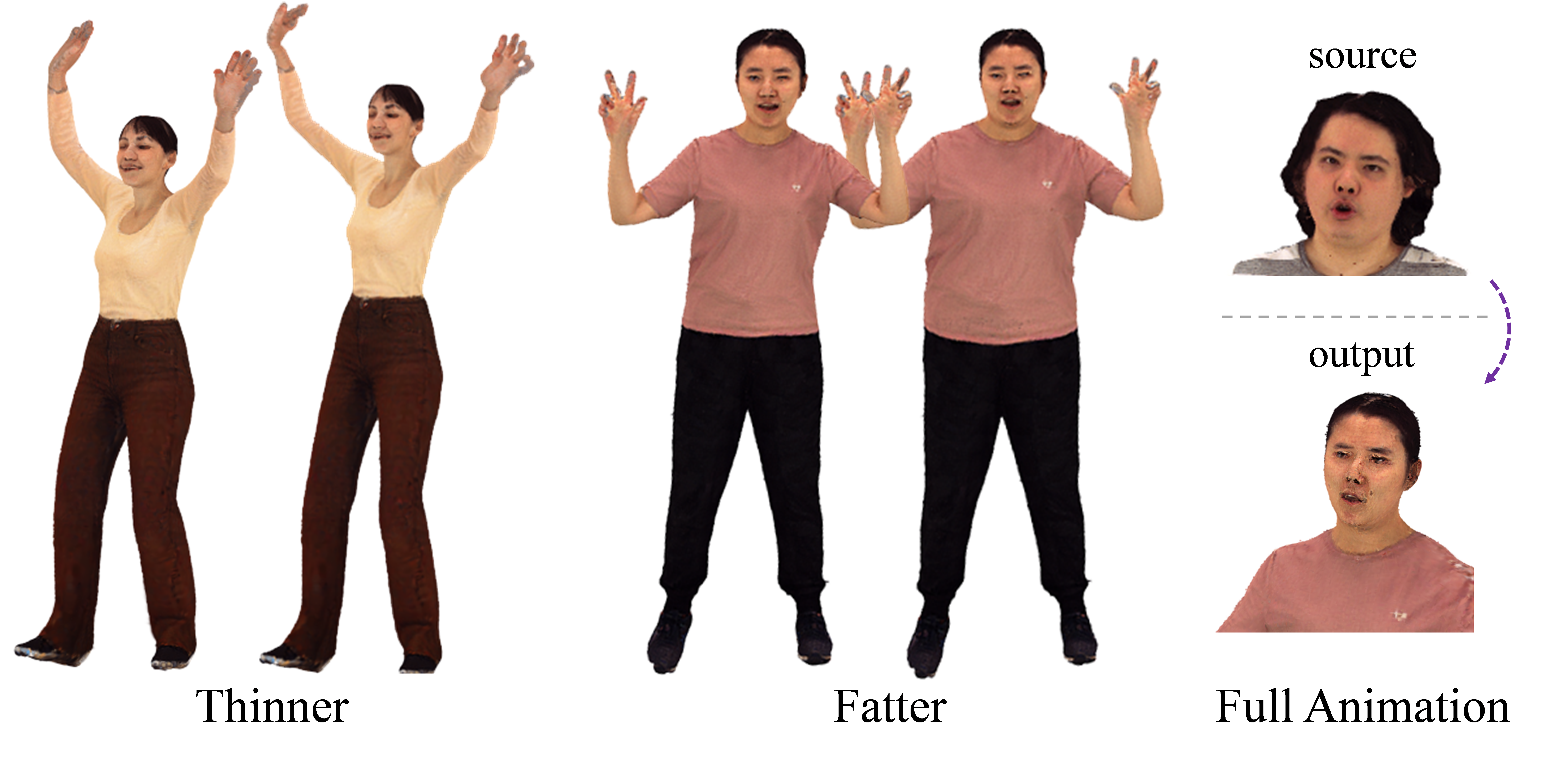}
        \caption{\textbf{Shape edit and full animation.} Our method easily modifies the model shape and facial expressions.}
        \label{fig:shape edit}
\end{figure}

Users can drive the Gaussian-based model to generate realistic results by controlling the underlying geometric structure. On the left of Fig.~\ref{fig:shape edit}, we adjust the shape using the first three principal components of the shape parameter, allowing the avatar to appear fatter and shorter or taller and thinner. Due to the comprehensive correspondence between the Gaussian fields and the underlying geometry in our design, the embedded facial Gaussian points can adapt to overall modifications, helping our model avoid artifacts during changes in body shape. Additionally, in the right of Fig.~\ref{fig:shape edit}, our model can realistically capture changes in facial expressions and hand poses. 

\subsection{Details of interactive interface} 
We built an interactive interface leveraging Sufel-rendering~\cite{ref13}, Nvdiffrast~\cite{ref57}, and drawing inspiration from~\cite{ref40}. The interface is divided into four functional modules: Render, Record, SMPL Transformation, and Edit. In the \textbf{Render} module, users can set up the camera view and adjust poses using data from the training dataset. It provides a convenient way to visualize the avatar under various configurations. \textbf{Record} module allows users to record animations by moving the camera and capturing the corresponding avatar movements. In the \textbf{SMPL Transformation} section, users can reset the avatar to a canonical pose, facilitating easy selection of specific regions and efficient texture projection. Additionally, users can modify the avatar's shape by adjusting the PCA parameters of the shape. This module also simplifies loading new poses to animate the model. The \textbf{Editing} module provides tools for selecting and modifying specific regions of the avatar. Users can choose regions using either a box selection or a pen selection tool and load the selected areas from a checkpoint. After selecting a region, users can edit its features through the interactive interface and preview changes in real time. Moreover, users can quickly load and project different textures from files onto the model. For a comprehensive overview of the interface and its capabilities, we strongly recommend watching the demo video.

\begin{figure}[hb]
    \centering
    \includegraphics[width=\linewidth]{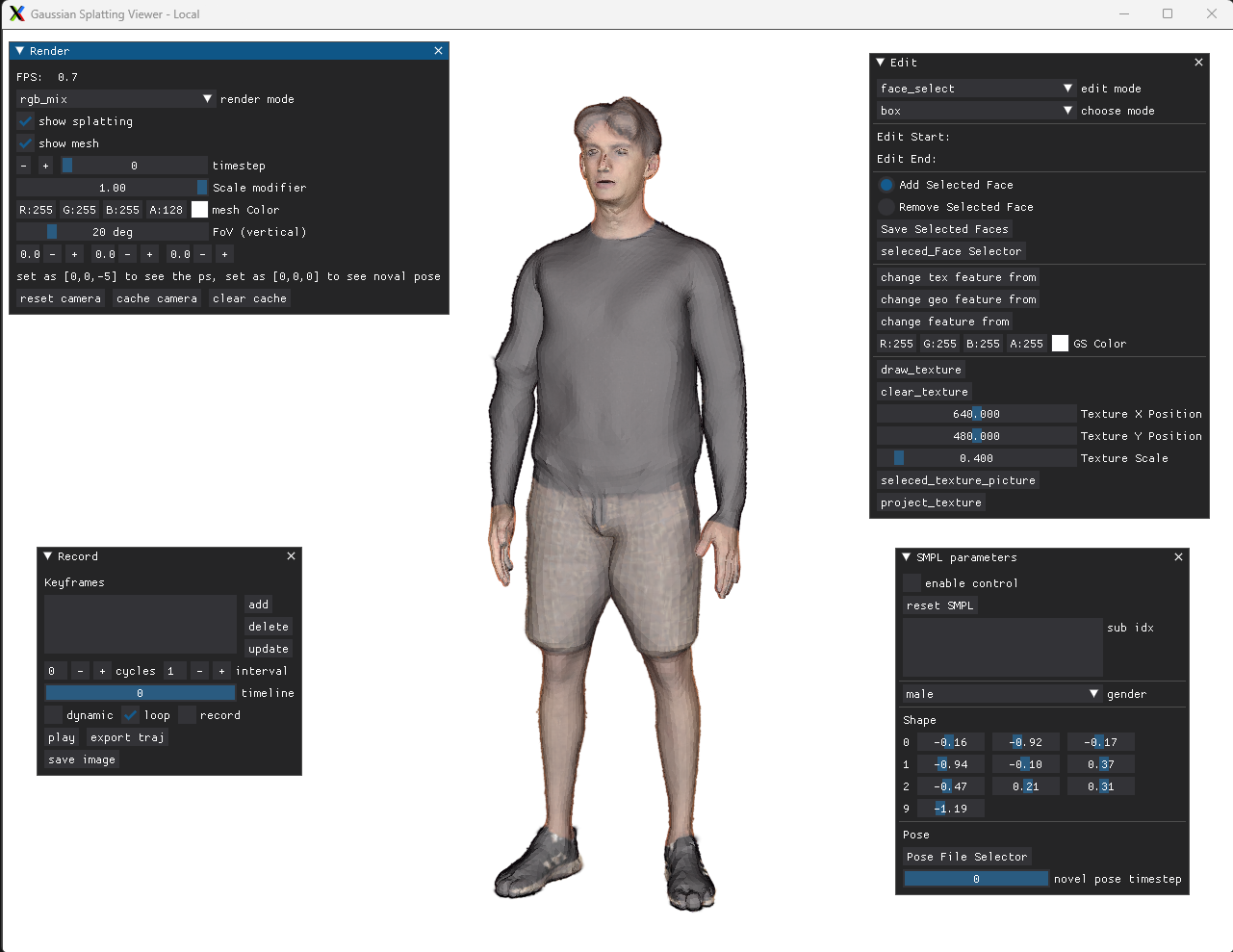}
    \caption{\textbf{Interactive interface}Our interface for AvatarBrush supports real-time rendering and editing, divided into four functional modules: Render, Record, SMPL Transformation, and Edit, enabling users to perform shape modification, pose control, texture stamping, and feature transfer.}
    \label{fig:interface}
\end{figure}

\section{Discussion and Limitation}
\subsection{Efficiency} In our framework, feature decoding is required only during the initial stage or after edits. For each frame, the process involves transforming the model and rendering it efficiently. With our interactive interface, rendering achieves approximately 30 FPS, while offline rendering reaches around 54 FPS. A novel identity model requires about 35 MB of storage, and training demands approximately 12 GB of memory on a single NVIDIA 4090 GPU. Compared to traditional Gaussian point cloud methods, our approach offers a significantly more compact way to store human models.

\begin{table}
\centering
\small
\begin{tabular}{lccc}
        \hline
              Model part & GS decoder & transform & render \\      
        \hline
             interface time(ms) &  17.19 & 8.29 & 12.39 \\
        \hline
        \label{table:inference}
    \end{tabular}
    \caption{\textbf{Rendering Efficiency.} Inference Time of each part in the model. }
\end{table}

\subsection{Challenges of Direct Gaussian Binding for Editing}
\begin{figure}[t]
    \centering
    \includegraphics[width=\linewidth]{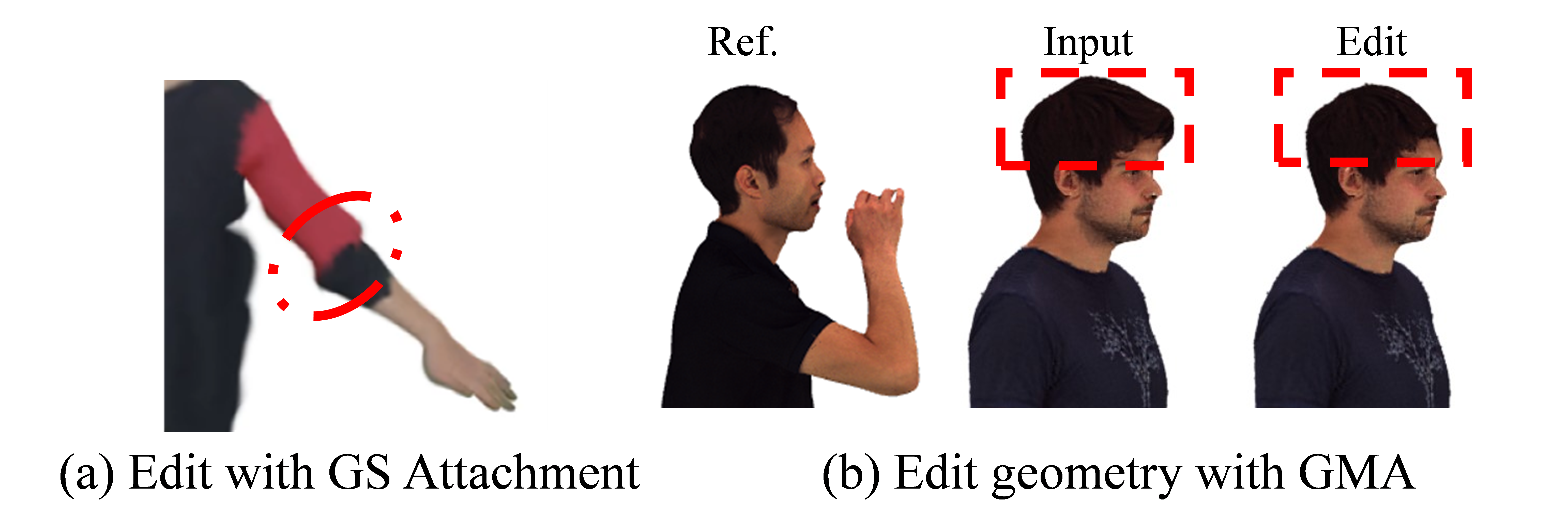}
    \caption{Compare with directly attaching the gaussian on the mesh. \textbf{(a) Edit with GS Attachment:} Direct attachment leads to geometric misalignment during editing. \textbf{(b) Edit geometry with GMA:} Our three-layer GMA representation enables coherent geometry edits via the feature layer.}
    \label{fig:edit}
\end{figure}

A critical design choice is the binding strategy between Gaussians and the mesh, which significantly impacts editability. We found that rigidly binding Gaussians to the local coordinates of the mesh severely complicates local editing. This tight coupling entangles geometry and appearance, meaning manual edits suffer from precision issues and seam artifacts. As shown in Fig.~\ref{fig:edit}, this can lead to geometric misalignment. To address this, our GMA representation introduces a three-layer decoupled structure. By performing edits on the high-level feature layer $f_{geo}, f_{tex}$ rather than the Gaussians themselves, our method avoids these artifacts, enabling robust and clean local modifications to both texture and geometry, as demonstrated in Fig.~\ref{fig:edit}.

\subsection{Limitation}
Although AvatarBrush achieves promising results in editable avatar reconstruction and local manipulation, several limitations remain that open avenues for future work.

\begin{figure}[t]
    \centering
    \includegraphics[width=\linewidth]{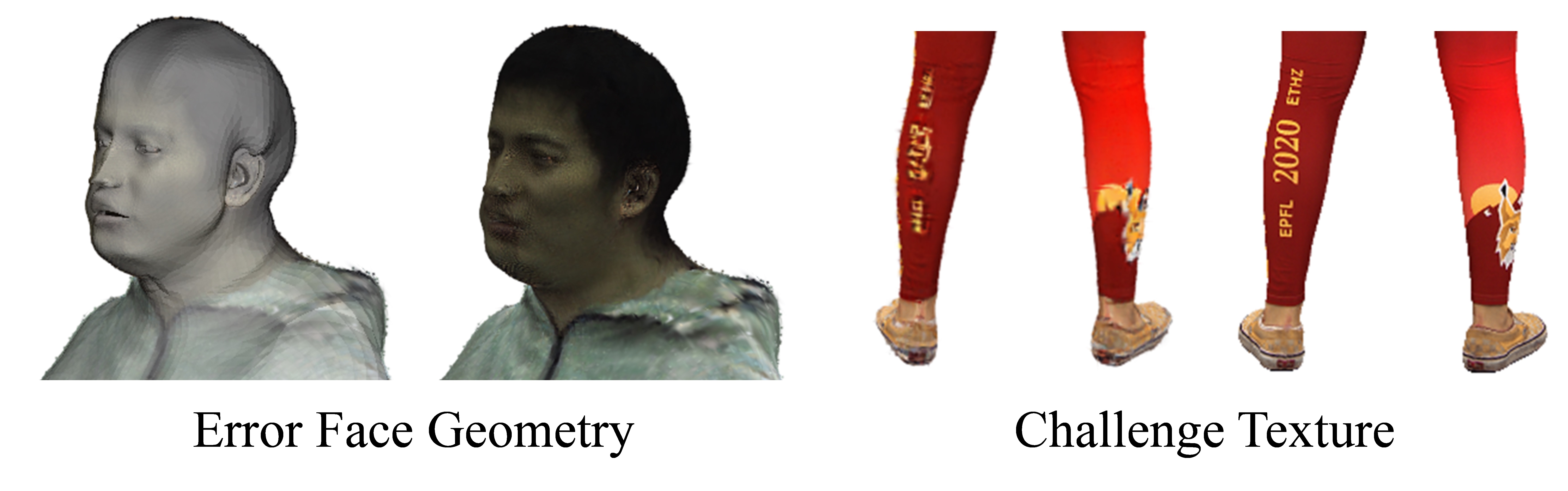}
    \caption{\textbf{Limitation.} Incorrect SMPL-X registration leads to overfitted facial geometry. The limited number of Gaussians in certain regions makes it challenging to generate high-frequency textures.}
    \label{fig:lim}
\end{figure}

\textbf{Dependency on SMPL-X Regression.}
Our current avatar morphing relies heavily on the accuracy of the underlying SMPL-X regression. In scenarios with strong occlusion or complex hand-object interactions, regression noise can propagate into the Gaussian layer. As illustrated in Fig.~\ref{fig:lim}(left), this noise can cause artifacts such as overfitted facial geometry, local misalignment, or texture stretching. Incorporating pose uncertainty estimation or learning-based deformation refinement could help mitigate these effects.

\textbf{Static Topology Assumption.}
Second, while our three-layer Gaussian representation establishes a strong correspondence between geometry and texture, it still assumes a relatively static topology. This design restricts the model’s ability to handle highly dynamic or topologically changing clothing, such as flowing skirts or loose hair. Extending our morphing mechanism to accommodate dynamic garments through physics-informed priors or explicit cloth simulation would further enhance realism.

\textbf{High-Frequency Texture Representation.}
Finally, the fidelity of fine details is constrained by the density of our Gaussian representation. As shown in Fig.~\ref{fig:lim}(right), the limited number of Gaussians allocated to certain regions makes it challenging to generate high-frequency textures, such as intricate patterns on clothing. Future work could explore adaptive Gaussian splitting mechanisms to better capture such details without significantly increasing computational cost.

Despite these limitations, our results demonstrate the effectiveness of Gaussian morphing for building editable and animatable avatars from monocular input. We believe that combining our representation with real-time tracking, dynamic lighting adaptation, and differentiable rendering will further expand its potential for digital human creation, virtual try-on, and immersive telepresence applications.


\section{Conclusion}
\label{sec:Conclusion}

In this paper, we introduce \textbf{AvatarBrush}, a novel and efficient framework for reconstructing an editable human avatar from monocular videos. Unlike previous approaches that rely on implicit representations or heavy volumetric optimization, our method enables intuitive local editing, fine-grained control, and efficient rendering, while maintaining high-fidelity geometry and texture.  

At the core of our framework lies the \textbf{Gaussian Morphing Avatar}, a new representation that establishes dense correspondences between 3D Gaussian primitives and a parametric human model through a carefully designed three-layer structure. This hierarchical design allows for clear disentanglement between geometry, texture, and appearance, enabling each to be optimized and manipulated independently. We further propose an \textbf{avatar morphing} mechanism that leverages the local properties of the parametric body to guide Gaussian deformation, ensuring semantic consistency and smooth transitions across poses and identities.  

Extensive experiments on multiple benchmark datasets demonstrate that AvatarBrush achieves superior quantitative results compared to state-of-the-art baselines, and produces visually coherent results under both novel-view and novel-pose settings. More importantly, our method supports flexible and user-friendly editing operations, such as garment transfer, texture stamping, and shape manipulation, without the need for retraining. These capabilities make AvatarBrush particularly well-suited for downstream applications in virtual try-on, digital human animation, and virtual production.  

In summary, AvatarBrush bridges the gap between high-quality avatar reconstruction and interactive controllability. By integrating Gaussian representations with parametric body priors, it provides a unified and extensible framework for building editable, animatable avatars from monocular inputs. In future work, we plan to explore the integration of dynamic cloth simulation, real-time motion retargeting, and cross-domain adaptation to further enhance realism and generalization. We believe that this work lays an important foundation for the next generation of controllable and expressive digital humans.

\begin{IEEEbiography}
[{\includegraphics[width=1in,height=1.25in, clip,keepaspectratio]{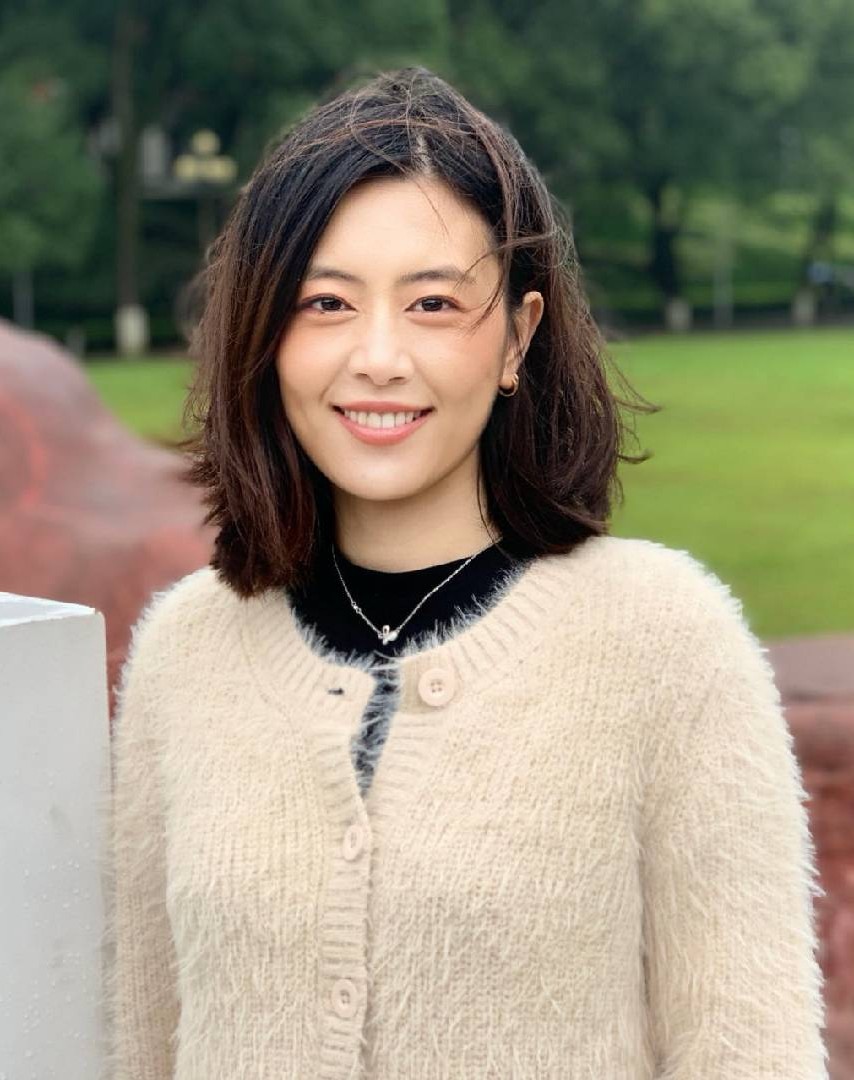}}]{Mengtian Li} currently holds a position as a Lecturer of Shanghai University, while simultaneously fulfilling the responsibilities of a Post-doc of Fudan University. She received a Ph.D. degree from East China Normal University, Shanghai, China, in 2022. She serves as a reviewer for CVPR, ICCV, ECCV, ICML, ICLR, NeurIPS, IEEE TIP, and PR, etc. Her research lies in 3D vision and computer graphics, focusing on human avatar animation and 3D scene understanding, reconstruction, and generation.
\end{IEEEbiography}

\vspace{-33pt}
\begin{IEEEbiography}
[{\includegraphics[width=1in,height=1.25in,clip,keepaspectratio]{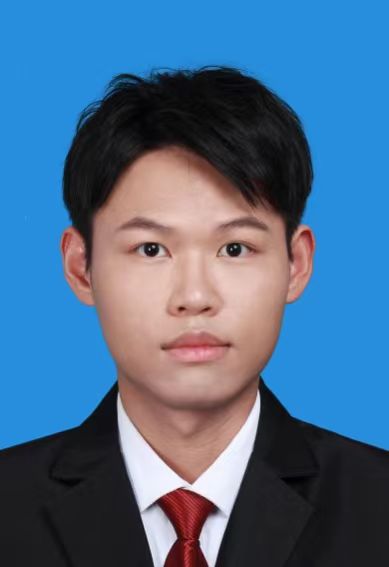}}]{Shengxiang Yao} received his Master's degree from the Shanghai Film Academy, part of Shanghai University. His research interests primarily focus on digital human reconstruction, with an emphasis on creating and editing avatars from video.
\end{IEEEbiography}
\vspace{-33pt}
\begin{IEEEbiography}[{\includegraphics[width=1in,height=1.25in,clip,keepaspectratio]{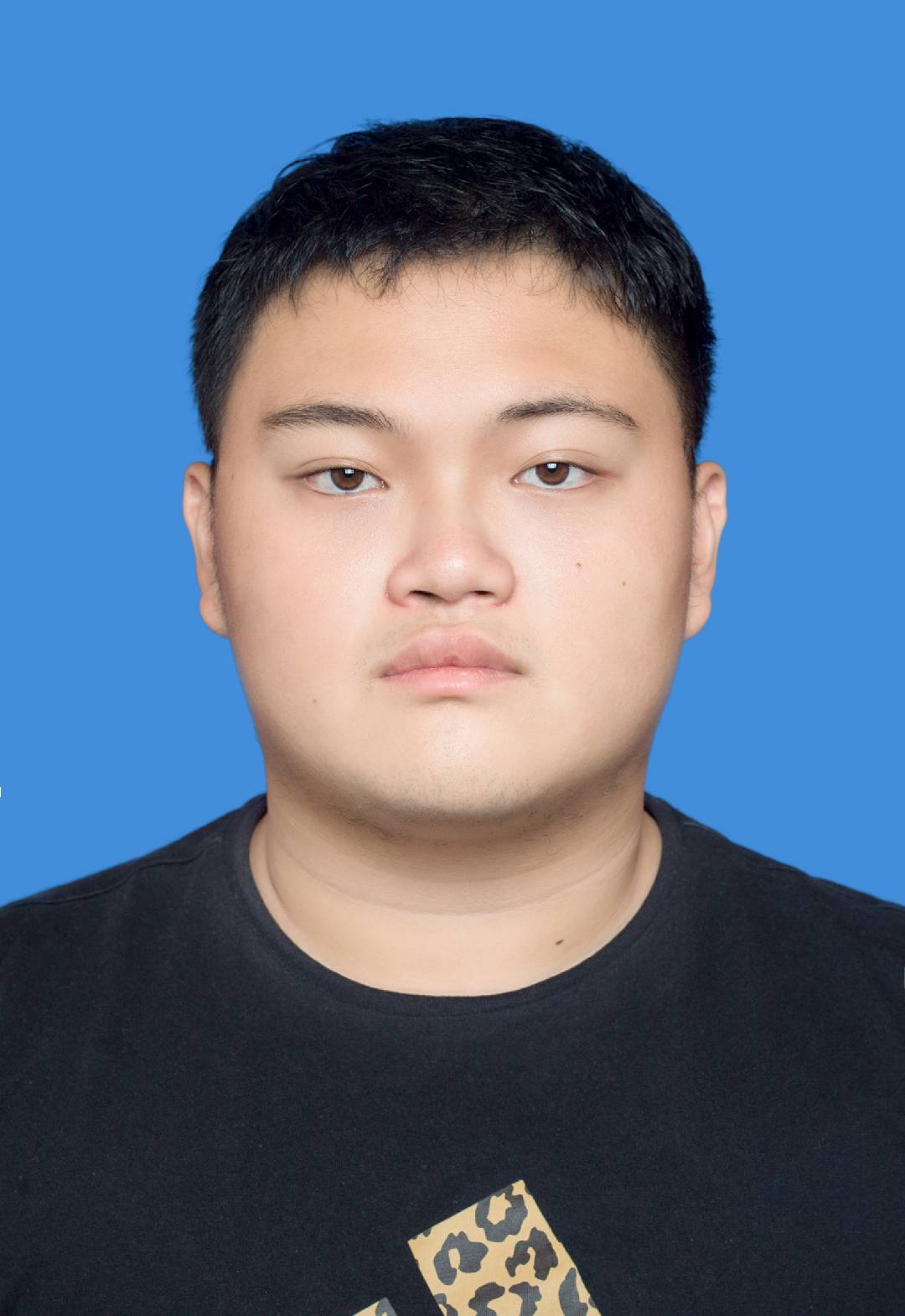}}]{Yichen Pan}
is currently pursuing an undergraduate degree in the Department of Digital Media Technology at Shanghai University. He has been a member of
Prof. Mengtian L’s lab since junior years. He has a strong interest in computer graphics, animation, and virtual human research. Since his undergraduate years, he has been actively involved in projects related to 3D reconstruction, motion generation, and digital avatar creation. His current research focuses on human motion generation, digital humans, and their applications in interactive media and virtual environments.
\end{IEEEbiography}

\begin{IEEEbiography}
[{\includegraphics[width=1in,height=1.25in,clip,keepaspectratio]{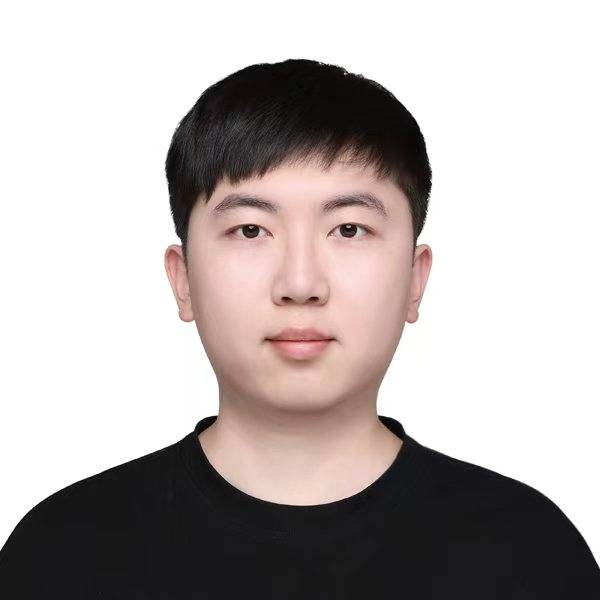}}]{Haiyao Xiao} is an AI Researcher at Tavus. Inc. He obtained his master’s degree from the University of Science and Technology of China in 2024. His research interests include 3D computer vision, neural rendering, and digital humans.
\end{IEEEbiography}
\vspace{-33pt}
\begin{IEEEbiography}[{\includegraphics[width=1in,height=1.25in,clip,keepaspectratio]{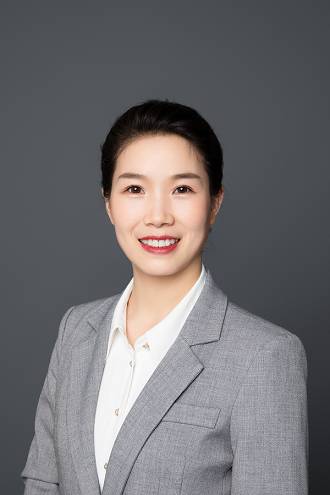}}]{Zhongmei Li} (Member, IEEE) is currently an associate professor at East China University of Science and Technology (ECUST), Shanghai, China. She received the B.S. degree in automation, and the M.S. and Ph.D. degrees in control science and engineering from Central South University, Changsha, China, in 2011, 2015, and 2019, respectively.
From September 2017 to September 2019, she was a Visiting Scholar at New York University, New York, NY, USA. Between January 2020 and July 2022, she worked as a Postdoctoral Fellow at the Key Laboratory of Smart Manufacturing in Energy Chemical Process, ECUST.
Her research interests include data-driven optimal control, complex industrial process control, and adaptive dynamic programming.
Dr. Li was the recipient of the 2023 IFAC Best Paper Prize and the 2020 IEEE CSS Beijing Chapter Young Author Prize.
\end{IEEEbiography}
\vspace{-33pt}
\begin{IEEEbiography}
[{\includegraphics[width=1in,height=1.25in,clip,keepaspectratio]{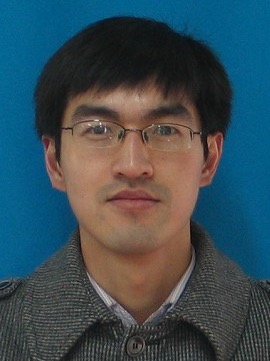}}]{Zhifeng Xie} received the Ph.D. degree in computer application technology from Shanghai Jiao Tong University, Shanghai, China. He was a Research Assistant with the City University of Hong Kong, Hong Kong. He is currently an Associate Professor with the Department of Film and Television Engineering, Shanghai University, Shanghai. He has published several works on CVPR, ECCV, IJCAI, IEEE Transactions on Image Processing, IEEE Transactions on Neural Networks and Learning Systems, and IEEE Transactions on Circuits and Systems for Video Technology. His current research interests include image/video processing and computer vision.
\end{IEEEbiography}
\vspace{-33pt}
\begin{IEEEbiography}[{\includegraphics[width=1in,height=1.25in,clip,keepaspectratio]{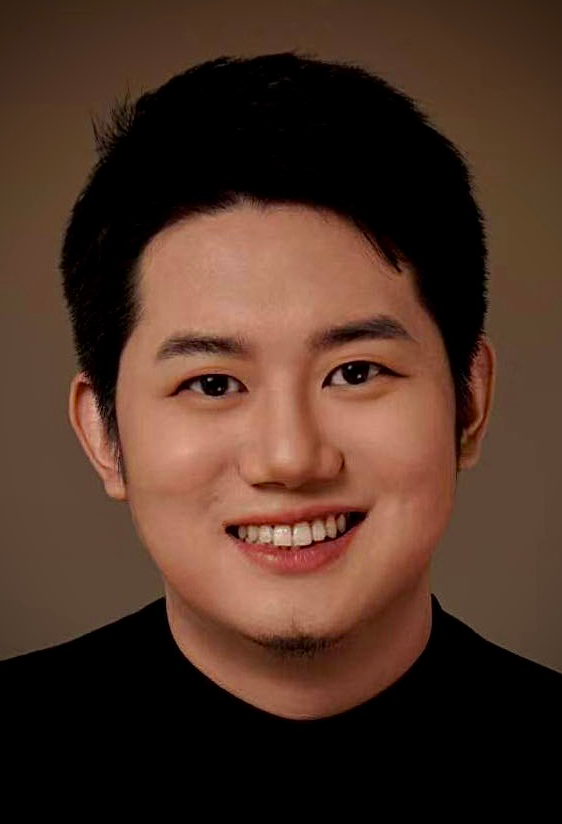}}]{Keyu Chen} is a senior AI researcher affiliated with Pinch Inc.. He obtained his master's and bachelor's degrees from the University of Science and Technology of China in 2021 and 2018. His research interests are mainly focused on digital human modeling, animation, and affective analysis.
\end{IEEEbiography}

\vspace{11pt}
\vfill
\end{document}